\documentclass[useAMS,usenatbib,usegraphicx]{mn2e}
\usepackage{natbib}
\usepackage{color}
\usepackage{hyperref} 
\hypersetup{colorlinks=true,citecolor=blue}
\usepackage[T1]{fontenc}
\usepackage{aecompl}
\def \bea {\begin{eqnarray}}
\def \ena {\end{eqnarray}}                  
\def \bee {\begin{equation}}
\def \ene {\end{equation}}
\def    \simlt  {\lower.5ex\hbox{$\; \buildrel < \over \sim \;$}}
\def    \simgt  {\lower.5ex\hbox{$\; \buildrel > \over \sim \;$}}

\def	\ltsim	{\simlt}

\newcommand     \mum    {\,\mu{\rm m}}  

\def	\cm		{\,{\rm {cm}}}

\def	\km		{\,{\rm {km}}}
\def	\B		{{\rm B}}

\def	\erg		{\,{\rm {ergs}}}

\def    \exp 		{\,{\rm {exp}}}
\def	\g		{\,{\rm g}}

\def	\K		{\,{\rm K}}

\def	\AU		{\,{\rm {AU}}}
\def	\pc		{\,{\rm {pc}}}

\def	\s		{\,{\rm s}}

\def    \ln  		{\,{\rm {ln}}}
\def    \yr  		{\,{\rm {yr}}}
\def	\erf		{\rm {erf}}

\def	\H		{\rm H}

\def	\xhat		{\hat{\bf x}}
\def	\yhat		{\hat{\bf y}}
\def	\zhat		{\hat{\bf z}}

\def	\rhat		{\hat{\bf r}}

\def	\hv			{{\bf h}}

\def	\V		{\rm V}

\def    \ma 		{\hat{\bf a}}

\def    \Bv     	{\bf  B}

\def    \rv     	{{\bf  r}}
\def    \kv     	{\bf  k}

\def	\Rv		{{\bf R}}

\def	\ba		{{\bf a}}
\def	\be		{{\bf e}}

\def	\bB		{{\bf B}}

\def 	\bE		{{\bf E}}

\def	\bH		{{\bf H}}

\def	\bJ		{{\bf J}}
\def	\bk		{\kv}
\def	\bM		{{\bf M}}

\def	\bP		{{\bf P}}
\def   	\bQ  		{{\bf Q}}

\def	\bp		{{\bf p}}

\def	\bv		{{\bf v}}

\def	\gas		{\rm {gas}}
\def	\th		{\rm {th}}

\def	\d		{\rm d}
\def	\rad		{\rm {rad}}
\def    \abs     	{\rm {abs}}
\def    \sca     	{\rm {sca}}

\def    \ext    	{\rm {ext}}
\def    \pol    	{\rm {pol}}
\def    \DG		  {\rm {DG}}
\def	\eff		{\rm {eff}}

\def	\drag		{\rm {drag}}

\def	\LTE		{\rm {LTE}}

\def    \coll        	{\rm {coll}}

\def	\ISRF		{\rm {ISRF}}

\def    \RAT		{\rm {RAT}}

\def	\hi		{\rm {highJ}}
\def	\lo		{\rm {lowJ}}

\def	\Bar		{\rm {Bar}}
\def	\nucl		{\rm {nucl}}

\def	\ali		{\rm {ali}}

\font\mib=cmmib10
\def\balpha{\hbox{\mib\char"0B}}
\def\bbeta{\hbox{\mib\char"0C}}
\def\bgamma{\hbox{\mib\char"0D}}
\def\bGamma{\hbox{\mib\char"00}}

\def\bdelta{\hbox{\mib\char"0E}}
\def\bepsilon{\hbox{\mib\char"0F}}
\def\bzeta{\hbox{\mib\char"10}}
\def\boldeta{\hbox{\mib\char"11}}
\def\btheta{\hbox{\mib\char"12}}
\def\biota{\hbox{\mib\char"13}}
\def\bkappa{\hbox{\mib\char"14}}
\def\blambda{\hbox{\mib\char"15}}
\def\bmu{\hbox{\mib\char"16}}
\def\bnu{\hbox{\mib\char"17}}
\def\bxi{\hbox{\mib\char"18}}
\def\bpi{\hbox{\mib\char"19}}
\def\brho{\hbox{\mib\char"1A}}
\def\bsigma{\hbox{\mib\char"1B}}
\def\btau{\hbox{\mib\char"1C}}
\def\bupsilon{\hbox{\mib\char"1D}}
\def\bphi{\hbox{\mib\char"1E}}
\def\bchi{\hbox{\mib\char"1F}}
\def\bpsi{\hbox{\mib\char"20}}
\def\bomega{\hbox{\mib\char"21}}

\title{Grain alignment by Radiative Torques in Special Conditions and Implications}

\author[Thiem Hoang \& A. Lazarian]{Thiem Hoang
$^1$\thanks{E-mail: hoang@cita.utoronto.ca}, and A. Lazarian
$^2$\thanks{E-mail: lazarian@astro.wisc.edu}\\ 
{$^1$ Canadian Institute for Theoretical Astrophysics, University of Toronto,
60 St. George Street, Toronto, ON M5S 3H8, Canada},\\
$^2$ Department of Astronomy, University of Wisconsin, Madison, WI 53706, USA}

\begin{document}
\maketitle
\begin{abstract}
Grain alignment by radiative torques (RATs) has been extensively studied for various environment conditions, including interstellar medium, dense molecular clouds, and accretion disks, thanks to significant progress in observational, theoretical and numerical studies. In this paper, we explore the alignment by RATs and provide quantitative predictions of dust polarization for a set of astrophysical environments that can be tested observationally. We first consider the alignment of grains in the local interstellar medium and compare predictions for linear polarization by aligned grains with recent observational data for nearby stars. We then revisit the problem of grain alignment in accretions disks by taking into account the dependence of RAT alignment efficiency on the anisotropic direction of radiation fields relative to magnetic fields. Moreover, we study the grain alignment in interplanetary medium, including diffuse Zodiacal cloud and cometary comae, and calculate the degree of circular polarization (CP) of scattered light arising from single scattering by aligned grains. We also discuss a new type of grain alignment, namely the alignment with respect to the ambient electric field instead of the alignment with the magnetic field. We show that this type of alignment can allow us to reproduce the systematic features of CP observed across a cometary coma. Our findings suggest that polarized Zodiacal dust emission may be an important polarized foreground component, which should be treated carefully in cosmic microwave background experiments.
\end{abstract}

\begin{keywords}
magnetic fields- polarization- dust, extinction
\end{keywords}

\section{Introduction}\label{sec:intro}
The polarization of starlight arising from differential extinction by nonspherical and aligned dust grains was discovered more than a half century ago (\citealt{Hall:1949p5890}; \citealt{Hiltner:1949p5851}). Interstellar dust grains are widely believed to be aligned with respect to magnetic fields, and it is essential to have a quantitative predictive theory of grain alignment if one wants to use the dust polarization as a reliable tracer of magnetic fields. Indeed, it is known observationally that the grain alignment changes with environments and sometimes fails.

The problem of grain alignment, however, proved to be one of the longest standing in astrophysics. In the process of research, a number of alignment mechanisms have been identified and quantified (see \citealt{2007JQSRT.106..225L} for a review), which substantially changed the initial paradigm of grain alignment based on the \cite{1951ApJ...114..206D} paramagnetic relaxation theory. Very importantly, an alignment mechanism based on radiative torques (RATs) acting on helical grains has become a favored mechanism to explain grain alignment. This mechanism was initially proposed by \cite{1976Ap&SS..43..291D},\footnote{\cite{Harwit:1970p6047} suggested a different radiative mechanism which is based on the interaction of an anisotropic radiation beam with different left- and right-handed photons with symmetric grains.} but was mostly ignored at the time of its introduction.\footnote{The mechanism was probably too radical for the community accustomed to the paramagnetic relaxation. It was also unclear why the alignment would favor observationally proven alignment of grains with long axes perpendicular to the magnetic field. The latter became clear only very recently with the advent of analytical theory of RAT alignment.} \cite{1996ApJ...470..551D} and \cite{1997ApJ...480..633D} (hereafter DW96 and DW97) reinvigorated the study of the RAT mechanism by numerically calculating RATs acting on grains of irregular shape.\footnote{It was essential that Bruce Draine has modified the publicly available DDSCAT code (\citealt{1994JOSAA..11.1491D}) to include RATs.} The strength of the torque made it impossible to ignore them, but the questions about the properties of RAT alignment for grains of different shapes remained. 

The quantitative study of the RAT alignment was presented first in \cite{2007MNRAS.378..910L} (henceforth LH07) where the analytical theory of RAT alignment was introduced and then elaborated in our series of papers (\citealt{Lazarian:2007p2442}; \citealt{2008ApJ...676L..25L}; \citealt{2008MNRAS.388..117H}, 2009ab). As a result of this work, it became clear why the grain alignment occurs with long axes perpendicular to the magnetic field, although the magnetic field just plays the axis of alignment. This work has opened an avenue for quantitative predictions of alignment for a variety of astrophysical situations. 

Historically, the interaction of the grain alignment research with observations was limited to accounting for the observed polarization. The goal was not very ambitious, namely, to find out a way to avoid gross contradictions of observations with plausible theoretical arguments. Nevertheless, this was not easy with the earlier theories. In comparison, the RAT mechanism seems to be able to address major puzzles presented by observations. For instance, the observation of polarized emission emanating from starless cores (see \citealt{WardThompson:2000p6330} and \citealt*{WardThompson:2002p6319}) initially seemed completely unexplainable.\footnote{These findings are also contrasting with observational claims based on visible and near-infrared radiation (\citealt{1995ApJ...448..748G}). The difference in results was explained in \cite{2007JQSRT.106..225L}.} Indeed, all mechanisms seemed to fail in such cores, which are presumably close to thermodynamic equilibrium (see \citealt{1997ApJ...490..273L}). The RATs seem to be too weak as well (DW96). However, \cite{2005ApJ...631..361C} (hereafter CL05) found that the efficiency of RATs increases fast with grain size (see also LH07), and thus large grains can still be aligned in dark clouds. They found that grains as large as $0.6 \mum$ can be aligned in dark clouds by the interstellar diffuse radiation attenuated by the column density with $A_{V} \approx 10$. This study proved the necessity of careful modeling of polarization if one is interested in magnetic fields in molecular clouds.

The pre-stellar cores studied in \cite{WardThompson:2000p6330} correspond to $A_{V} = 50-60$; the shielding column, assuming uniformity, is approximately half of these values. However, \cite{Crutcher:2004p6727} pointed out that polarization data do not sample the innermost core regions,\footnote{The peaks $A_{V}$ of 150 were claimed for the clouds in \cite{Pagani:2004p6728}.  These peaks are likely not to produce polarized dust emission.} and this provides an explanation for polarization ``holes'' (see \citealt{2000ApJ...531..868M};\citealt{Lai:2002p6336};
\citealt{2002ApJ...574..822M}). The reported decrease in the percentage polarization with the optical depth also agrees well with the findings in CL05.

The approach of CL05 was further elaborated in the studies by \cite{2007ApJ...663.1055B} and \cite*{2009A&A...502..833P}, in which the synthetic maps obtained via magnetohydrodynamic (MHD) simulations were analyzed. In particular, \cite{2007ApJ...663.1055B} calculated the actual value of anisotropy degree $\gamma_{\rad}$, mean intensity $J_{\lambda}$ of radiation field, and grain temperature inside a turbulent molecular cloud. The study confirmed the ability of using aligned grains to trace magnetic fields in dense clouds and proved a decrease of percentage polarization at the highest $A_{V}$. Note that these works were in contrast to the previous studies, which used rather arbitrary criteria (e.g., $A_{V}=3$) for the alignment to shut down, or even more unrealistic assumption that all grains were perfectly aligned. 

Before proceeding with detailed modeling of dust polarization, which is the main subject of this paper, let us outline a few major features of the RAT quantitative theory that the modeling is based upon. In LH07, we subjected to scrutinize the properties of RATs. Using a simple analytical model (AMO) of a helical grain we studied the basic properties of RATs and the alignment driven by such RATs. The analytical results were found to be in good correspondence with numerical calculations for irregular grains obtained with DDSCAT (\citealt{1994JOSAA..11.1491D}). Evoking the generic properties of the RAT components, we explained the RAT alignment of grains both in the absence and presence of magnetic fields. Intentionally, for the sake of simplicity, in LH07, we studied a simplified dynamical model to demonstrate the effect of the RAT alignment. This model disregarded the wobbling of grain axes with respect to the angular momentum that arises from thermal fluctuations (\citealt{1994MNRAS.268..713L}; \citealt{1997ApJ...484..230L}) and thermal flipping of grains (\citealt{1999ApJ...516L..37L}; \citealt{1999ApJ...520L..67L}). These simplifications allowed us to provide the first quantitative model of the RAT alignment, which, however, required further elaboration. In particular, we found that RAT alignment in general occurs with attractor points of high angular momentum (i.e., $J> J_{\th}$ with $J_{\th}$ being the thermal angular momentum, hereafter high-$J$) and of low angular momentum ($J\sim J_{\th}$, hereafter low-$J$). We obtained criteria for the grains to be aligned with high-$J$ and low-$J$ attractor points as a function of the angle between the radiation anisotropy direction and magnetic fields and the ratio of the magnitude of the RAT components.

Using AMO, \cite{2008MNRAS.388..117H} studied the RAT alignment by taking into account the grain wobbling, thermal flipping, and random collisions of gas atoms. We found that the random collisions of gas atoms, which act to randomize the orientation of grains when they are thermally rotating, can enhance the RAT alignment when the grains are aligned with high-$J$ attractor points. We also found that for superparamagnetic grains, the presence of high-$J$ attractor points occurs more frequently (\citealt{2008ApJ...676L..25L}) due to the joint action of paramagnetic relaxation and RATs. In that sense, the inclusion of iron atoms into dust grains can increase the fraction of grains aligned with high-$J$ attractor points and the degree of alignment eventually. However, a comprehensive study on the degree of alignment as a function of radiation field, magnetic field and dust physics is necessary for modeling dust polarization.

All earlier works assumed perfect alignment for the grains larger than a critical size, which is drawn using the condition of suprathermal rotation induced by RATs (CL05; \citealt{2007ApJ...663.1055B}; \citealt*{Pelkonen:2007p6094}). In addition, the grain rotation induced by RATs was obtained assuming that the radiation anisotropy direction is parallel to the magnetic field direction. As a result, the earlier results seem to overestimate the degree of grain alignment and ultimate polarization level. Indeed, \cite{2009ApJ...697.1316H} showed that the rate of suprathermal rotation is maximum when the radiation anisotropy direction is parallel to the magnetic field and decreases with the increasing angle $\psi$ between the two directions, provided that the magnetic field always is the axis of alignment. Since the polarization is determined by the fraction of grains with the suprathermal rotation, such a decrease indicates that the alignment efficiency tends to decrease with the increasing $\psi$. This important prediction of RAT alignment was confirmed by \cite{2011A&A...534A..19A}.

The efficiency of the Purcell torques, including torques due to H$_{2}$ formation, photoemission, and the variation of the accommodation coefficient over the grain surface (\citealt{1979ApJ...231..404P}), which are fixed within the grain body, decreases when the grain rapidly wobbles. The effect of thermal wobbling was quantified in \cite{1997ApJ...484..230L} and the results of this study were used by \cite{1999ApJ...516L..37L} and \cite{1999ApJ...520L..67L} (hereafter LD99ab) to predict new effects of grain dynamics, namely, thermal flipping and thermal trapping. The thermal flipping is a phenomenon in which the wobbling angle between the grain symmetry axis and the angular momentum overcomes the separatrix angle of $90^{\circ}$. When this occurs, the torques that are fixed in the grain body change their direction with respect the angular momentum. When the thermal flipping occurs fast enough, the Purcell torques get averaged out and the grain rotates thermally in spite of the presence of the uncompensated pinwheel torques, i.e., it is thermally trapped. 

Recently, the picture of thermal flipping was challenged by \cite{Weingartner:2009p5709}. He found that the grain does not experience thermal flipping as a result of internal relaxation, instead, it tends to be frozen at the separatrix (i.e., when the grain symmetry axis becomes perpendicular to the angular momentum). This is due to the fact that the diffusion coefficient of internal relaxation vanishes when the grain approaches the separatrix. \cite{2009ApJ...695.1457H} improved the treatment of thermal flipping in LD99ab by considering the realistic situation in which the grain is subject not only to internal relaxation, but also to random bombardment by gas atoms. The latter is subdominant in most of the time, but it becomes important when the grain approaches the separatrix. As a result, random collisions prevent the grain from being frozen at the separatrix and therefore realistic grains do flip unlike the claim in \cite{Weingartner:2009p5709}. Using the improved treatment of thermal flipping, \cite{2009ApJ...695.1457H} showed that the H$_{2}$ formation torques can help to increase the alignment arising from RATs through the additional spin-up.

Observational evidence for RAT alignment are numerous and increasingly available (\citealt{2008ApJ...674..304W}; \citealt{2010ApJ...720.1045A}; \citealt{2011PASJ...63L..43M}; \citealt{2011A&A...534A..19A}; \citealt{2012A&A...541A..52V}). In particular, fundamental features of RAT alignment, such as the dependence of alignment on radiation intensity and anisotropy direction, have been tested and confirmed by observations (see \cite{2010ApJ...720.1045A}; \citealt{2011A&A...534A..19A}). Very recently, \cite{Andersson:2013gw} report an important evidence for the enhancement of grain alignment due to $\H_{2}$ formation torques in the reflection nebula IC 63. The detailed modeling for RAT alignment in the presence of H$_{2}$ formation torques by Hoang, Lazarian \& Andersson (2014, in preparation) shows that the enhancement of alignment for some lines of sight can originate from the local enhancement of H$_{2}$ torques within the IC 63 nebula, but the RAT alignment induced by strong radiation from the $\gamma$ Cas star still plays a leading role.

This paper is intended to provide a detailed modeling of grain alignment and dust polarization using the latest theoretical progress. In particular, we are going to explore the grain alignment in special environments for which observational tests are feasible, such as the local interstellar medium (LISM), Zodiacal cloud (ZC), cometary comae, as well as accretion disks. Special attention is paid to understanding the main alignment mechanisms in these environments and implications for probe of interplanetary magnetic fields (IMFs) and for polarized cosmic microwave background (CMB) studies. 

The paper is structured as follows. In \S \ref{sec:alignmech}, we summarize the principal processes and their characteristic timescales involved in grain alignment. A description of basic features of grain alignment by RATs and their degree of alignment is presented in \S \ref{sec21}. In \S \ref{sec:mod} we present a method for modeling of linear polarization by aligned grains in a given molecular cloud. The possibility of grain alignment by RATs for accretion disks is revisited in \S \ref{sec:accdisk}. We apply our method to model grain alignment and calculate the linear polarization for the LISM in \S \ref{sec:LISM}. The alignment of grains in ZC and cometary comae is discussed in \S \ref{sec:cirpol} and \ref{sec:comet}, where predictions of degree of circular polarization (CP) arising from scattering by aligned grains are presented. The discussion and summary is presented in \S \ref{sec:dis} and \ref{sec:sum}.

\section{Grain alignment mechanisms and characteristic timescales}\label{sec:alignmech}
In this section, we summarize the principal processes involved in grain alignment and their characteristic timescales.
In general, the alignment process of dust grains can be divided into two stages: (i) the alignment of grain axes with the angular momentum $\bJ$ and (ii) the alignment of $\bJ$ with the ambient magnetic field $\Bv$. The former usually occurs over a short timescale due to internal relaxation processes, namely, inelastic relaxation, Barnett relaxation and nuclear relaxation (see Section \ref{sec:Bar}). The latter occurs over a much longer timescale and involves a number of competing processes including the rotational damping, the Larmor precession of grain magnetic moment around the ambient magnetic field, and the precession time of grain electric dipole moment around the electric field, and the precession of grain around the radiation direction.

\subsection{Grain geometry}
In the following, we consider oblate spheroidal grains with moments of inertia $I_{1}>I_{2}=I_{3}$ along the grain's principal axes. Let us denote $I_{\|}=I_{1}$ and $I_{\perp}=I_{2}=I_{3}$. They take the following forms: 
\bea
I_{\|}=\frac{2}{5}Ma^{2}=\frac{8\pi}{15}\rho ba^{4},\\
I_{\perp}=\frac{4\pi}{15}\rho a^{2}b\left(a^{2}+b^{2}\right),
\ena
where $a$ and $b$ are the lengths of semi-major and semi-minor axes of the oblate spheroid with axial ratio $r=a/b>1$, and $\rho$ is the grain material density. The factor $h=I_{\|}/I_{\perp}$ is then equal to
\bea
h=\frac{2a^{2}}{a^{2}+b^{2}}=\frac{2}{1+s^{2}},\label{eq:h}
\ena
where $s=b/a\equiv 1/r<1$.

\subsection{Barnett relaxation}\label{sec:Bar}
\cite{Barnett:1915p6353} pointed out that a rotating paramagnetic body gets magnetized with the magnetic moment parallel to the angular velocity.\footnote{This is an inverse of the Einstein-de Haas effect, that was used to measure the spin of the electron. To grain alignment theory the Barnett effect was introduced by \cite{1976Ap&SS..43..291D}, who noticed that the effect should induce the magnetic moment of grains.} This effect can easily be understood using a classical model. 

Consider a paramagnetic grain which is rotating with the angular velocity ${\Omega}$. As the grain rotates with angular velocity ${\Omega}$, the torque acting on the electron spin is $\frac{\d}{dt} {\bJ}={\Omega}\times {\bJ}$. The equivalent torque can be induced by a magnetic field ${\bH}_{\B}$ acting on the magnetic moment $\bmu$ associated with the spin, i.e. $1/c {\bmu}\times {\bH}_{\B}$, which provides the Barnett-equivalent field $H_{\rm eqv}=\left(J c/{\mu}\right) \Omega$. The latter can also be presented as 
\bea
{\bH}_{\textrm{eqv}}=\frac{\Omega}{\gamma_{g}},\label{eq:315}
\ena
where  $\gamma_{g}=\mu/(Jc)=eJ/(m_{e}Jc)=e/(m_{e}c)$, with $e$ the electron charge and $m_{e}$ the electron mass, is the magneto-mechanical ratio of an electron. 

\cite{1979ApJ...231..404P} showed that as the grain wobbles, the changes of magnetization in the grain axes cause the internal relaxation, which he termed "Barnett relaxation". \cite{1999ApJ...520L..67L} (LD99a) revisited the problem by taking into account both spin-lattice and spin-spin relaxation (see \citealt{Morrish:1980}). 
The Barnett relaxation time is equal to (\citealt{1979ApJ...231..404P})
\bea
\tau_{\Bar}=\frac{\gamma_{g}^{2}I_{\|}^{3}}{VK(\omega)h^{2}(h-1)J^{2}},\label{eq:tauBar}
\ena
where $V$ is the grain volume, $h=I_{\|}/I_{\perp}$, $K(\omega)=\chi''(\omega)/\omega_{1}$ with $\omega_{1}=(h-1)J\cos\theta/I_{\|}$ and $\theta$ the angle between $\bJ$ and $\ba_{1}$ (LD99a).

For oblate spheroidal grains, we obtain
\bea
\tau_{\Bar}\approx 0.5\hat{\rho}^{2}a_{-5}^{7}\hat{s}\left(\frac{1+\hat{s}^{2}}{2}\right)^{2}\left(\frac{J_{\d}}{J}\right)^{2}
\left[1+\left(\frac{\omega_{1}\tau_{el}}{2}\right)^{2}\right]^{2}\yr,~~~\label{eq:316}
\ena
where $a_{-5}=a/10^{-5}\cm$ with $a$ being the grain size, $\hat{s}=s/0.5$, $\hat{\rho}=\rho/3\g\cm^{-3}$ with $\rho$ being the grain material density, $\tau_{el}\sim \tau_{2}\sim 2.9\times 10^{-12}f_{p}^{-1}$s with assumption of $f_{p}=0.1$ is the spin-spin coupling time; $J_{\d}=\sqrt{I_{\|}k_{\B}T_{\d}/(h-1)}$ is the dust thermal angular momentum.\footnote{The relaxation of electron spins results from the spin-lattice and spin-spin relaxation, with time scales $\tau_{1}\gg\tau_{2}$, so here we adopted $\tau_{el}\sim \tau_{2}$ (\citealt{Draine:1996p6977}).}

Although \cite{1979ApJ...231..404P} considered grains with both electron and nuclear spins, his study missed the effect of internal relaxation related to nuclear spins. LD99a found that for astrophysical grains of realistic composition nuclear spins induce a new type of relaxation, which was termed "nuclear relaxation" by LD99a. This relaxation can be understood in a simple-minded approach in terms of much stronger \textit{equivalent} magnetic field given by Equation~(\ref{eq:315}). Indeed, this field is proportional to the mass of the species involved. As the paramagnetic relaxation is proportional to $H^2_{\rm eqv}\chi'' \propto\gamma_{g}^{2}\chi(0)\tau\propto m^{2}(1/m^{2})\tau\sim \tau$, and as $\tau_{n}\gg\tau_{el}$, one can understand the nature of the dominance of the nuclear relaxation, which has a characteristic time:
\bea
\tau_{\nucl}&\approx& 3.1\times 10^{-6} \hat{\rho}^{2}a_{-5}^{7}\hat{s}\left(\frac{1+\hat{s}^{2}}{2}\right)^{2}\left(\frac{J_{\d}}{J}\right)^{2}\nonumber\\
&&\times \left[1+\frac{(\omega_{1}\tau_{n})^{2}}{2}\right]^{2}\yr,~~~\label{tau_nucl}
\ena
where $\tau_{n}$ is the relaxation rate induced by the nucleus-nucleus and electron-nucleus spin interactions (see LD99a). 

\subsection{Rotational damping}
The damping of grain rotation mainly arises from collisions with gas atoms and emission of infrared photons by the grain (\citealt{Purcell:1971p2747}; \citealt{1993ApJ...418..287R}).

Collisions of a grain with gas atoms consist of sticking collisions in which gas atoms stick to the grain surface followed by their evaporation. In the grain frame, the mean torque arising from the sticking collisions for an axisymmetric grain rotating around its symmetry axis tends to zero when averaged over the grain revolving surface. On the other hand, the evaporation induces a non-zero mean torque, which is parallel to the rotation axis (see \citealt{1993ApJ...418..287R}; \citealt{Lazarian:1997p5348}).

The rotational damping rate due to the dust-gas collisions is given by
\bea
\frac{\langle \Delta J\rangle}{\Delta t}=-\frac{J}{\tau_{\gas}},
\ena
where $\tau_{\gas}$ is the gaseous damping time:
\bea
\tau_{\gas}&=&\frac{3}{4\sqrt{\pi}}\frac{I_{\|}}{n_{\H}m_{\H}
v_{\th}a^{4}\Gamma_{\|}},\nonumber\\
&=&7.3\times 10^{4} \hat{\rho}\hat{s}a_{-5}\left(\frac{100\K}{T_{\gas}}\right)^{1/2}\left(\frac{30
\cm^{-3}}{n_{\H}}\right)\left(\frac{1}{\Gamma_{\|}}\right) \yr,~~~~\label{eq:taugas}
\ena
where $v_{\th}=\left(2k_{\B}T_{\gas}/m_{\H}\right)^{1/2}$ is the thermal velocity of a gas atom of mass $m_{\H}$ in a plasma with temperature $T_{\gas}$ and density $n_{\H}$. Above, $\Gamma_{\|}$ is a geometrical parameter, which is equal to unity for spherical grains. This timescale is comparable to the time required for the grain to collide with an amount of gas equal to its own mass.

IR photons emitted by the grain carry away part of the grain's angular momentum, resulting in damping of the grain rotation. The rotational damping rate by IR emission can be written as
\bea
\tau_{\rm IR}^{-1}=F_{\rm IR} \tau_{\gas}^{-1},\label{eq:tauIR}
\ena
where $F_{\rm IR}$ is the rotational damping coefficient for a grain having an equilibrium temperature $T_{\d}$ (see \citealt{1998ApJ...508..157D}), which is given by
\bea
F_{\rm IR}	=\left(\frac{0.91}{a_{-5}}\right)\left(\frac{u_{\rad}}{u_{\ISRF}}\right)^{2/3}
\left(\frac{30 \cm^{-3}}{n_{\H}}\right)\left(\frac{100 \K}{T_{\gas}}\right)^{1/2},\label{eq:FIR}
\ena
where $u_{\rad}$ is the energy density of the interstellar radiation field (ISRF) and $u_{\rm ISRF}=8.64\times 10^{-13}\erg\cm^{-3}$ is the energy density of the local ISRF as given by \cite*{Mezger:1982p4014}.

The rotational damping rate is then given by
\bea
\tau_{\rm drag}^{-1}=\tau_{\gas}^{-1}+\tau_{\rm IR}^{-1}=\tau_{\gas}^{-1}\left(1+F_{\rm IR}\right).\label{eq:taudamp}
\ena

For large grains of size $a>10^{-5}\cm$, the gaseous damping is dominant and $\tau_{\rm drag}\approx \tau_{\gas}$. For small grains of $a=10^{-6}-10^{-5}\cm$, the damping by IR emission becomes dominant for most of the interstellar medium (ISM), except for molecular clouds (\citealt{1998ApJ...508..157D}). In the above equation, the damping due to passing ions (plasma drag) and ion collisions is disregarded due to their negligible importance for $a>0.01\mum$ grains considered in this paper.

Usually, we represent the grain angular momentum and timescales in units of the thermal angular momentum $J_{\th}$ and gaseous damping time $\tau_{\gas}$. The former is given by
\bea
J_{\th}&=&\sqrt{I_{\|}k_{\B}T_{\gas}}=\sqrt{\frac{8\pi\rho a^{5}s}{15}k_{\B}
T_{\gas}},\nonumber\\
&=&5.9\times 10^{-20}\hat{s}^{1/2}\hat{\rho}^{1/2}
 a_{-5}^{5/2}\left(\frac{T_{\gas}}{100\K}\right)^{1/2} {\g\cm}^{2}{\rad 
\s}^{-1}.~~~\label{eq:Jth}
\ena
Similarly, the thermal angular velocity is equal to
\bea
\omega_{\th}&=&\left(\frac{2k_{\B}T_{\gas}}{I_{\|}}\right)^{1/2},\nonumber\\
&=&3.3\times10^{5}\hat{s}^{1/2}\hat{\rho}^{-1/2}a_{-5}^{-5/2}\left(\frac{T_{\gas}}{100\K}\right)^{1/2}\s
.\label{eq:ome_th}
\ena
 
\subsection{Larmor precession of $\bJ$ around $\Bv$}
A rotating grain can acquire a magnetic moment thanks to the Barnett effect (see \citealt{Barnett:1915p6353}; \citealt{1969mech.book.....L}) and the rotation of its charged body (\citealt{1972MNRAS.158...63M}; \citealt{1976Ap&SS..43..257D}). The Barnett effect, which is shown to be much stronger than the latter, induces a magnetic moment proportional to the grain angular velocity:
\bea
\bmu_{\Bar}=\frac{\chi(0)V\hbar}{g\mu_{B}}\bomega,\label{eq:muBar}
\ena
where $g$ is the gyromagnetic ratio, which is $\approx 2$ for electrons, and $\mu_{B}=e\hbar/2m_{e}c$ is the Bohr magneton (see \citealt{Draine:1996p6977} and references therein).

The interaction of this magnetic moment with the external static magnetic field, governed by the torque $[-\bmu_{\Bar}\times \bB]=-\mu_{\Bar}B\sin\xi\hat{\phi}\equiv I_{\|}\omega\sin\xi d\phi/dt \hat{\phi}$, causes the regular precession of the grain angular momentum around the magnetic field direction. The rate of such a Larmor precession denoted by $\tau_{B}$, is given by
\bea
\tau_{B}&=&\frac{2\pi}{d\phi/dt}=\frac{2\pi I_{1}\omega}{\mu_{\Bar}B},\nonumber\\
&=&0.84\hat{\rho}^{-1/2}\hat{\chi}^{-1}\hat{B}^{-1} a_{-5}^{2}~\yr,
\label{eq:tauB}
\ena
where $f_{p}=0.01$ is assumed, and $\hat{B}=B/5\mu$G and $\hat{\chi}=\chi(0)/10^{-4}$ are the normalized magnetic field and magnetic susceptibility, respectively.

\subsection{Precession $\ma_{1}$ around the anisotropic direction $\kv$}
\cite{2007MNRAS.378..910L} found that the third component of RAT efficiency, $Q_{e3}$, induces the grain precession around the anisotropic direction of radiation field $\kv$. The timescale for such a RAT precession is defined by
\bea 
\tau_{k}&=&\frac{2\pi}{|d\phi/dt|},\nonumber\\
&=& 3.2\times10^{3}
\hat{\rho}^{1/2}\hat{T}^{1/2}\left(\hat{\lambda}\hat{u}_{\rad}\right)^{-1}\hat{Q}_{e3}^{-1}a_{-5}^{1/2}
\yr,\label{eq:tauk}
\ena 
where $\hat{T}=T_{\gas}/100\K$ and
\bea 
\frac{d\phi}{dt}=\frac{\gamma_{\rad} u_{\rad}\lambda a_{\eff}^{2}}{2I_{1}\omega\sin\xi}{\bf
Q}_{\Gamma}.\hat{\Phi}
=\frac{\gamma_{\rad}u_{\rad}\lambda a_{\eff}^{2}}{I_{1}\omega} Q_{e3}.\label{eq79}
\ena 
Above, $\hat{Q}_{e3}=Q_{e3}/10^{-2}$, $\hat{\lambda}=\lambda/1.2 \mum$, $\hat{u}_{\rad}=u_{\rad}/u_{\rm ISRF}$, and $\gamma_{\rad}=0.1$ for the anisotropy of radiation field. In deriving Equation (\ref{eq:tauk}) $\omega=\omega_{\th}$ has been used. For axisymmetric grains, the two first components of RAT efficiency, $Q_{e1}$ and $Q_{e2}$ are equal to zero, while the third component $Q_{e3}$ is non-zero (see \citealt{2007MNRAS.378..910L}). Therefore, $Q_{e3}$ produces the fast precession of grains around ${\kv}$. 

The alignment of grains whether with the radiation direction or magnetic field depends on the rate of grain precession around these axes. From Equations (\ref{eq:tauk}) and (\ref{eq:tauB}), the ratio of the precession rate around the radiation to that around the magnetic field is equal to
\bea  
\frac{\tau_{k}}{\tau_{B}}&=3.3\times 
10^{3}\hat{\rho}\hat{\chi}\hat{T}^{1/2}a^{-3/2}_{-5}\left(\frac{\hat{B}}
{\hat{\lambda}\hat{u}_{\rad}}\right)\hat{Q}_{e3}^{-1}.\label{eq:taukB}
\ena

It can be checked easily that for the typical diffuse ISM, $\tau_{B}/\tau_{k} \sim 10^{-3}$, i.e., the Larmor precession is much faster than the precession induced by RATs. Therefore, the magnetic field plays the role of the alignment axis.

\subsection{Precession of grain electric dipole moment $\bp$ around $\bE$}

In addition to magnetic moments, dust grains posses electric dipole moments. The electric dipole moment of a grain consists of the intrinsic moment due to molecules and substructures with polar bonds, and the moment due to the asymmetric distribution of grain charge. It can be written as
\bea
p^{2}= (\epsilon Qa)^{2}+p_{\rm int}^{2},\label{eq:ptot}
\ena
where $Q$ is the grain charge and $\epsilon a$ is the displacement between the grain charge centroid and the center of mass (see \citealt{1998ApJ...508..157D}). The grain charge centroid is present for irregular grains even if they perfectly conducting (\citealt{1975duun.book..155P}).

For ultrasmall grains (e.g., polycyclic aromatic hydrocarbons), the intrinsic dipole moment $p_{\rm int}$ dominates, but for larger grains the dipole moment due to the charge distribution becomes dominant. The latter can be rewritten as
\bea
p= 1.0\times 10^{-15} \left(\frac{U}{0.3\rm V}\right)\left(\frac{\epsilon}{10^{-2}}\right)a_{-5}~ {\rm statC} \cm,\label{eq:pE}
\ena
where $U\approx Q/a$ is the grain electrostatic potential.

In an ambient electric field $\bE$, the grain electric dipole moment precesses around $\bE$ at a rate
\bea
\Omega_{E}\equiv\frac{2\pi}{\tau_{E}}=\frac{\bp.\bE}{J},\label{eq:omegaE}
\ena
where $\tau_{E}=2\pi/|d\phi/dt|$ with $\phi$ being the precession angle of $\bJ$ around $\bE$. Here, we have assumed that the dipole moment ${\bf p}$ is coupled to the grain angular momentum.

The electric field can be produced by static charges and/or the relative motion of charged grains across magnetic fields. The latter is given by $c\bE=-\bv\times \Bv$, which can be significant for grains of supersonic motion.
Assuming the electric field $E =10^{-5} \V\cm^{-1}=10^{-5}\times 10^{8}/c~ {\rm statV cm^{-1}}$ and the grain electrostatic potential $U=0.3 \V$ for Equation (\ref{eq:omegaE}),\footnote{The electric field in the earth atmosphere is estimated as $E=6\times 10^{-1} \V\cm^{-1}$ at the height of $2\km$ and decreases with the increasing latitude.} we obtain
\bea
\tau_{E}&=&7.0 \times 10^{-4} \hat{\rho}\left(\frac{\epsilon}{10^{-2}}\right)
\left(\frac{U}{0.3 \rm V}\right)^{-1}\left(\frac{E}{10^{-5}\rm V\cm^{-1}}\right)^{-1}\nonumber\\
&\times&\left(\frac{\omega}{\omega_{\th}}\right)
\left(\frac{T_{\gas}}{100\K}\right)^{1/2}a_{-5}^{1/2}\yr.~~~~ \label{eq:tauE}
\ena

\subsection{Davis-Greenstein Paramagnetic Alignment}
A classical mechanism of grain alignment based on paramagnetic dissipation was proposed by \cite{1951ApJ...114..206D}. The underlying idea of the mechanism is that, a paramagnetic grain gets magnetized with an instantaneous magnetization $\bM$ parallel to the induced magnetic field $\Bv$.
For a rotating grain, the continuous magnetization induces the dissipation of grain rotational energy into vibrational energy,  which brings the grain into alignment with $\bJ$ parallel to $\Bv$.  

Due to the paramagnetic dissipation, the angle $\beta$ between $\bJ$ and $\Bv$ gradually decreases with time as
\bea
I_{\|}\omega\frac{d\beta}{dt}=-K(\omega)VB^{2}\omega\sin\beta\cos\beta,
\ena
where $V$ is the grain volume and $K(\omega)=\chi''(\omega)/\omega$.

The above equation can be rewritten as
\bea
\frac{d\beta}{dt}=-\frac{\sin\beta\cos\beta}{\tau_{\rm DG}},\label{eq:dbeta_tauDG}
\ena
where the characteristic timescale $\tau_{\rm DG}$ is given by
\bea
\tau_{\rm DG}&=&\frac{I_{\|}}{K(\omega)VB^{2}}\equiv\frac{2\rho a^{2}}{5K(\omega)B^{2}}\nonumber\\
&\approx& 1.2\times 10^{6}\hat{\rho}\hat{T}_{d}\hat{B}^{-2}a_{-5}^{2}
\left(\frac{1.2\times10^{-13}\s}{K(\omega)}\right)
\yr,~~~\label{eq:tau_DG}
\ena
where $\hat{T}_{d}=T_{d}/15\K$ with $T_{d}$ being the dust temperature. For slowly rotating grains with $\omega < 10^{8} \s^{-1}$, $K(\omega) \approx 1.2\times 10^{-13}\s$, and $K(\omega)$ decreases rapidly with the increasing $\omega$ (see \citealt{Draine:1996p6977}).

\subsection{Mechanical alignment}

Mechanical alignment mechanism proposed by \cite{Gold:1952p5849} is based on impact of atoms and molecules to spin the grain up. Basically, each impact deposits an increment of angular momentum $\delta J=am_{\H}v_{\rm flow}$ to the grain. Assuming random walk for the impacts, the total increase of grain rotational energy per unit of time is equal to $(\Delta J)^2/\Delta t=R_{\coll}(\delta J)^{2}$, where $R_{\coll}=n_{\H}v_{\rm flow}\pi a^{2}$
is the rate of collisions. Following \cite{1976Ap&SS..43..291D}, the timescale for the mechanical alignment is defined as the time required to spin up the grain from an initial angular momentum $J_{0}$:
\bea
\tau_{\rm Gold}=\frac{J_{0}^{2}}{(\Delta J)^{2}/(\Delta t)}
=\frac{J_{0}^{2}}{\gamma n_{\H}m_{\H}^{2}v_{\rm flow}^{3}\pi a^{4}}
\ena

Assuming $J_{0}=J_{\th}$, the alignment by the gaseous flow occurs over a timescale
\bea
\tau_{\rm Gold}&=&\frac{16\rho k_{\B} T_{\gas} a}{15\gamma_{\rm flow}^{2}n_{\H} m_{\H} ^{2}v_{\rm flow}^{3}} \nonumber\\
&=&1.7\times10^{7}\hat{\gamma}a_{-5}\left(\frac{T_{\gas}}{100\K}\right)
\left(\frac{n_{\H}}{30\cm^{-3}}\right)^{-1}\nonumber\\
&&\times\left(\frac{v_{\rm flow}}{10^{5}\cm\s^{-1}}\right)^{-3}~~~\yr,\label{eq:tauGold}
\ena
where $\hat{\gamma}=\gamma_{\rm flow}/0.1$ is the anisotropy of the gaseous flow, and
$v_{\rm flow}$ is the velocity of the flow relative to the ambient gas and for a spherical of equivalent radius $a$. Several mechanisms of dust acceleration based on interactions of MHD turbulence with charged grains have been shown to drive grains to supersonic motion (\citealt{2003ApJ...592L..33Y}; \citealt*{2004ApJ...616..895Y}; \citealt*{2012ApJ...747...54H}). The mechanical alignment for helical grains of subsonic motion was discussed in \cite{Lazarian:2007p2442}. 

\section{Grain alignment by Radiative Torques}\label{sec21}
Consider a grain subject to an external regular torque $\bGamma$, and a damping torque arising from collisions with gas atoms and the emission of IR photons. The evolution of the grain angular momentum is then governed by the conventional equation of motion:
\bea
\frac{d\bJ}{dt}=\bGamma-\frac{\bJ}{\tau_{\rm drag}},\label{eq:dJdt}
\ena
where $\tau_{\rm drag}$ is the rotational damping time given by Equation (\ref{eq:taudamp}), and the second order effect of random collisions by gas atoms is disregarded.

The value of grain angular momentum in a stationary state, denoted by $J_{\max}$, can be obtained by setting $d\bJ/dt=0$. Thus,
\bea
J_{\max}=\Gamma_{J}\times \tau_{\rm drag},\label{eq:Jmax}
\ena
where $\Gamma_{J}$ is the the torque component projected onto the direction of $\bJ$.

\subsection{Anisotropic Radiative Torques}\label{sec23}
Let $u_{\lambda}$ be the energy density of radiation field per wavelength $\lambda$ and $\gamma_{\rad}$ its anisotropy. Denote $u_{\rad}=\int u_{\lambda}d\lambda$, which is the energy density of radiation. The magnitude of RAT arising from the interaction of radiation field with an irregular grain of size $a$ is given by
\bea
\Gamma_{\lambda}=\gamma_{\rad} \pi a^{2}
u_{\lambda} \left(\frac{\lambda}{2\pi}\right)Q_{\Gamma},\label{eq:GammaRAT}
\ena
where $Q_{\Gamma}$ is the RAT efficiency (see DW96).

For the case the radiation direction parallel to the axis of maximum moment of inertia, $\ba_{1}$, LH07 found that the RAT efficiency can be approximated by a power law:
\bea
Q_{\Gamma}\approx 0.4\left(\frac{{\lambda}}{a}\right)^{\eta},\label{eq:QAMO}
\ena
where $\eta=0$ for $\lambda \ltsim 2a$  and $\eta=-3$ for $\lambda \gg a$. 

From Equations (\ref{eq:Jmax}) and (\ref{eq:QAMO}) one can determine the maximum angular momentum induced by RATs as:
\bea
\frac{J_{\max}^{\RAT}}{J_{\th}}&=&\left(\int \Gamma_{\lambda} d\lambda\right) \frac{\tau_{\rm drag}}{J_{\th}},\\
&\approx &200\hat{\gamma}_{\rad}\hat{\rho}^{1/2}a_{-5}^{1/2}
\left(\frac{30\cm^{-3}}{n_{\H}}\right)\left(\frac{100\K}{T_{\gas}}\right)\nonumber\\
&&\times
\left(\frac{\bar{\lambda}}
{1.2\mum}\right)\left(\frac{u_{\rad}}{u_{\ISRF}}\right)\left(\frac{\overline{Q_{\Gamma}}}{10^{-2}}\right)
\left(\frac{1}{1+F_{\rm IR}}\right),~~~~~\label{eq:Jmax_RAT}
\ena
where $\hat{\gamma}_{\rad}=\gamma_{\rad}/0.1$,
\bea
\bar{\lambda}&=&\frac{\int \lambda u_{\lambda} d\lambda}{u_{\rad}},\label{eq:wavemean}\\
\overline{Q}_{\Gamma}&=&\frac{\int Q_{\Gamma} \lambda u_{\lambda}d\lambda}{\overline{\lambda}u_{\rad}},\label{eq:Qmean}
\ena
are the wavelength and RAT efficiency averaged over the entire radiation field spectrum, respectively.

Using Equations (\ref{eq:QAMO})-(\ref{eq:Qmean}), we can calculate the maximum rotation rate $J_{\max}^{\RAT}$ due to RATs for a grain of size $a$ embedded in the radiation field $u_{\lambda}$.

The characteristic timescale for RATs to spin a grain from thermal rotation to suprathermal rotation is defined as
\bea
\tau_{\rm spin-up}&=&\frac{J_{\th}}{dJ/dt}=\frac{J_{\th}}{\Gamma-J/\tau_{\rm drag}}\nonumber\\
&=&\frac{\tau_{\rm drag}}{J_{\max}/J_{\th}-1},\label{eq:tau_spinup}
\ena
where Equation (\ref{eq:dJdt}) has been used.

\subsection{Dependence of RAT alignment on radiation direction}
The maximum grain angular momentum induced by RATs, $J_{\max}^{\RAT}$ (Eq. \ref{eq:Jmax_RAT}), is obtained assuming that the anisotropic direction of the radiation field $\kv$ is parallel to the axis of major inertia $\ba_{1}$ (i.e., $\Theta=0$). In the presence of a magnetic field, the grain usually rotates about the axis of alignment, $\Bv$. Thus, the exact value $J_{\max}^{\RAT}$ may be reduced due to projection effect.

\cite{2009ApJ...697.1316H} found that $J_{\max}^{\RAT}$ decreases with the increasing angle $\psi$ between $\kv$ and $\Bv$. Since the RAT alignment tends to occur with $\bJ$ parallel to $\Bv$, only the RAT component projected onto $\Bv$ is to spin grains up to the maximum angular momentum:
\bea
J_{\max}^{\RAT}(\psi)=J_{\max}^{\RAT}(\psi=0)\cos\psi,\label{eq:Jmaxpsi}
\ena
where $J_{\max}^{\RAT}$ is given by Equation (\ref{eq:Jmax_RAT}). The above equation is easily obtained by using Equation (\ref{eq:Jmax}) and $\cos\Theta=1$ for the perfect internal alignment. Indeed, the projection of RATs onto $\Bv$ is equal to $\Gamma_{J}\propto Q_{J}(\psi)=Q_{e1}\cos\Theta\cos\psi-Q_{e2}\sin\Theta\cos\psi$. For $\cos\Theta=1$, we obtain $Q_{J}(\psi)= Q_{e1}\cos\psi=Q_{J}(\psi=0)\cos\psi$ or $\Gamma_{J}(\psi)=\Gamma_{J}(0)\cos\psi$.

From Equation (\ref{eq:Jmaxpsi}) we can see that $J_{\max}^{\RAT}(\psi=90^{\circ})=0$. However, this value is obtained for the case without internal thermal fluctuations. When such thermal fluctuations are taken into account, it is expected that $J_{\max}^{\RAT}(\psi=90^{\circ}) \sim J_{d}$ (i.e., grains rotate thermally regardless of radiation intensity; \citealt{2008MNRAS.388..117H}).

In addition to dependence on $\psi$, the existence of high-$J$ attractor points depend on other parameters, including grain shape, size, and spectrum of the radiation field (see \citealt{2007MNRAS.378..910L}).

\subsection{Suprathermal rotation and critical size of aligned grains}\label{sec24}

In the RAT alignment mechanism, some grains are aligned with high-$J$ attractor points with $J_{\hi}=J_{\max}(\psi)$ given by Equation (\ref{eq:Jmax_RAT}), whereas most grains are driven to low-$J$ attractor points having $J_{\lo}\sim J_{\th}$. The alignment of grains with high-$J$ attractor points is stable if grains rotate suprathermally, i.e., $J_{\max}(\psi)\gg J_{\th}$. Using the Langevin equations to follow the RAT alignment of grains in the presence of gas collisions, \cite{2008MNRAS.388..117H} found that grains can have a stable alignment when $J_{\max}(\psi)/J_{\th}\approx 3$. 

We assume that the critical size of suprathermal rotation is identical to the critical size of aligned grains, denoted by $a_{\ali}$. Therefore, throughout this paper, we are interested only in $a_{\ali}$. Taking the usage of the condition for the stable alignment $J_{\max}(\psi)\approx 3J_{\th}$ for Equation (\ref{eq:Jmax_RAT}), one can determine $a_{\ali}$ as a function of the environment parameters, including $u_{\rad}, n_{\H}$, and $T_{\gas}$. Since RATs increase rapidly with $a$, grains larger than $a_{\ali}$ would suprathermally be rotating, i.e., $J_{\max}(\psi)> 3J_{\th}$, which results in the perfect internal alignment of grain axes with the angular momentum.

Using the degree of alignment as a function of grain size inferred from the best-fit model to the observational data, Hoang, Lazarian, \& Martin 2014 (submitted to ApJ) find that the alignment of $a>a_{\ali}$ grains plays a crucial role for the optical and IR polarization, whereas the alignment of $a<a_{\ali}$ grains can contribute to the polarization in ultraviolet (UV). Therefore, the determination of $a_{\ali}$ is very important for modeling dust polarization.

\subsection{Degree of RAT alignment}\label{sec25}
First, we consider the alignment of grains larger than $a_{\ali}$, which have suprathermal rotation induced by RATs. Let $Q_{X}=\langle G_{X}\rangle$ with $G_{X}= \left(3\cos^{2}\theta-1\right)/2$ be the alignment efficiency of the grain axis of major inertia with angular momentum, and let $Q_{J}=\langle G_{J}\rangle$ with $G_{J}= \left(3\cos^{2}\xi-1\right)/2$ be the alignment efficiency of the angular momentum and the magnetic field. Here the angle brackets denote the average over the ensemble of grains. Let $f_{\hi}$ be the fraction of grains that are aligned with high-$J$ attractors. Since grains that are aligned with high$-J$ attractors have perfect alignment, we can write the alignment degree of angular momentum as
\bea
Q_{J}=f_{\hi}+(1-f_{\hi})Q_{J,\lo}.\label{eq:QJ}
\ena
where $Q_{J,\lo}$ is the degree of alignment of grains with low-$J$ attractors.

LH07 showed that $f_{\hi}$ in general depends on a number of parameters, including the grain size $a$, the ratio of RAT efficiency $q^{\max}$, the radiation direction $\psi$, and the radiation field $u_{\rad}$. In this paper, we won't attempt to compute an exact value of $f_{\hi}$, which appears impossible given its dependence on numerous uncertain parameters. Thus, we treat $f_{\hi}$ as a model parameter throughout this paper.

The Rayleigh reduction factor is defined as
\bea
R(a)=<G_{J}G_{X}>\approx f_{\hi}+(1-f_{\hi})Q_{J,\lo}Q_{X,\lo},~~~\label{eq:Rayleigh}
\ena
where we have used the fact that the alignment of suprathermally rotating grains with high-$J$ attractors (i.e, $J^{2}\gg I_{\|}kT_{\rm d}$) corresponds to the perfect internal alignment, i.e., $Q_{X,\hi}=1$. Equation (\ref{eq:Rayleigh}) can be simplified further using the approximation $Q_{J,\lo}\approx 1$. Here, we disregard the correlation of $Q_{X}$ and $Q_{J}$, which is minor for suprathermal grains (see \citealt{1999MNRAS.305..615R}; Hoang et al. 2014).

The degree of internal alignment for the low-$J$ attractor point $Q_{X,\lo}$ at which the grain axes undergo strong thermal fluctuations due to the exchange of vibrational and rotational energy can be calculated by
\bea
Q_{X,\lo}=\int_{0}^{\pi} G_{X}f_{\rm VRE}(\theta, J_{\lo})\sin\theta d \theta,~~~~\label{eq:QXl}
\ena
where
\bea
f_{\rm VRE}(\theta, J_{\lo})=Z\exp\left(-\frac{J_{\lo}^{2}}{2I_{\|}k_{\B}T_{\d}}\left[1+(h-1)\sin^{2}\theta\right]\right)~~~
\ena
is the distribution function of grain axis with respect to angular momentum with $Z$ being the normalization determined by $\int_{0}^{\pi} f_{\rm VRE} \sin\theta d\theta=1$, and $h=I_{\|}/I_{\perp}$ (see \citealt{1997ApJ...484..230L}). \cite{2008MNRAS.388..117H} found that $J_{\lo}$ is determined by the internal thermal fluctuations, which is comparable to vibrational energy, i.e., $J_{\lo}\simeq \left(2I_{\|}k_{\B}T_{\d}\right)^{1/2}$.

Small grains with $a\le a_{\ali}$ can be weakly aligned by paramagnetic relaxation and are important for the polarization at UV wavelength as well as spinning dust polarization  (\citealt{Hoang:2013tp}), whereas the $a>a_{\ali}$ aligned grains dominate the starlight polarization in optical and IR (Hoang et al. 2014). Therefore, it is reasonable to disregard the contribution of small aligned grains while dealing with the optical/IR polarization and set $R(a<a_{\ali})=0$.

One important parameter in polarization modeling is the fraction of grains aligned with high-$J$ attractor points, $f_{\hi}$. For ordinary paramagnetic material, the RAT alignment is rarely perfect, i.e., $f_{\hi}<1$. $f_{\hi}$ is significantly increased for grains with superparamagnetic inclusion (\citealt{2008ApJ...676L..25L}). For alignment with high-$J$ attractor points, $f_{\hi}$ is able to reach unity (i.e., perfect alignment) when the collisional excitation arising from gas bombardment is taken into account (\citealt{2008MNRAS.388..117H}).

\section{Modeling Dust Polarization}\label{sec:mod}
Below, we describe our general approach to model the alignment of grains by RATs induced by radiation fields in a molecular cloud and calculate linear polarization of starlight by the aligned grains. 
\subsection{Radiative transfer}\label{sec32}
\begin{figure}
\includegraphics[width=0.45\textwidth]{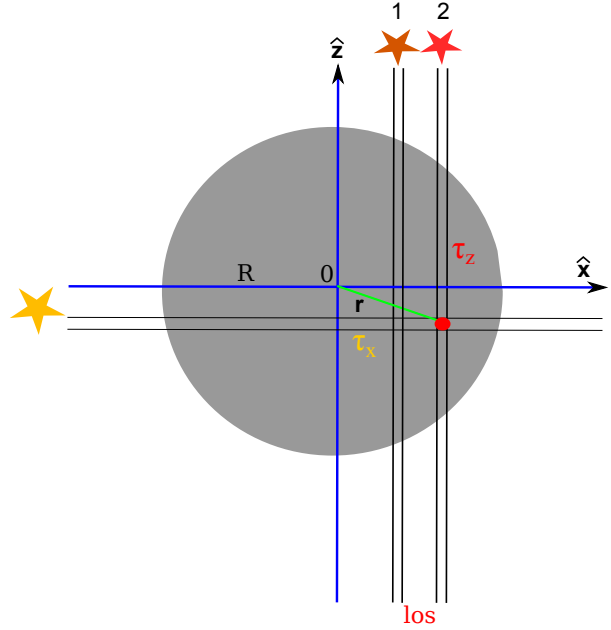}
\caption{Schematic illustration of a molecular cloud illuminated by a nearby star. The plane containing lines of sight $\xhat\zhat$ is perpendicular to the sky plane. The star is assumed to be located in the sky plane perpendicular to the plane $\xhat\zhat$ and illuminates grains in the $\xhat$ direction. Two background stars (1 and 2) are observed through the cloud with different values of optical depth, $\tau_{z}$.}
\label{fig:f1}
\end{figure}

We consider a molecular cloud in which grains are illuminated both by the stellar radiation from a nearby star as well as attenuated ISRF. The distance from the cloud to the star $d_{\star}$ is assumed to be much larger than the cloud radius. We assume that the ISRF is anisotropic with the degree of anisotropy $\eta$, whereas the stellar radiation is completely anisotropic (parallel beam). A schematic illustration of our study is shown in Figure \ref{fig:f1}.

Let $\tau_{x}$ and $\tau_{z}$ be the optical depths due to dust extinction along the $\xhat$ and $\zhat$ directions, respectively. We divide the plane of sightlines into a two dimension grid of $N_{x}\times N_{z}$ cells. The gas density and density of radiation energy at each cell are given by $n_{\H}(x,z)$ and $u_{\lambda}(x,z)$, respectively. Also, the gas temperature $T_{\gas}$ in general is a function of $x$ and $z$.

Provided the temperature of the star $T_{\star}$ and the distance $d_{\star}$ from the cloud, we can derive the spectral energy density $u_{\lambda}$ as follows:
\bea
u_{\lambda}=\frac{4\pi}{c}\frac{\int I_{\lambda}d\Omega}{4\pi}, \label{eq:uwav}
\ena
where $I_{\lambda}$ is the intensity of radiation and $\Omega$ is the solid angle in steradians (see Appendix \ref{apdx:a}). Due to the dust extinction, the energy density decreases as
\bea
u_{\lambda}=\frac{4\pi B_{\lambda}(T_{\star})}{c}\left(\frac{R_{\star}}{d_{\star}}\right)^{2}e^{-\tau_{x}(\lambda)}
+u_{\lambda}(\rm ISRF),\label{eq:urad}
\ena
where the second term denotes the attenuated ISRF inside the cloud. Here, the IR emission of dust grains is disregarded because the RATs induced by the IR emission are negligible for interstellar grains (i.e., $a\ll \lambda$).

\begin{figure}
\includegraphics[width=0.45\textwidth]{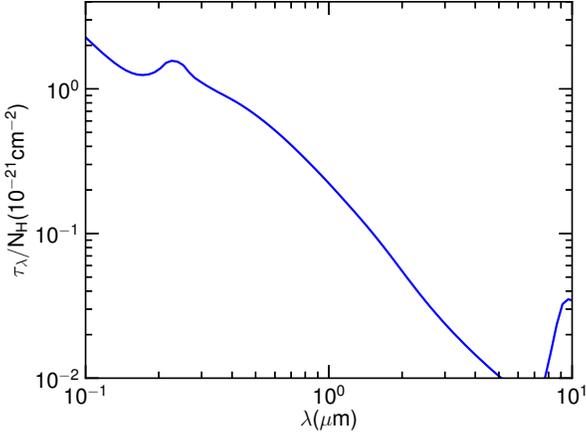}
\caption{Optical depth per column density N$_{\H}$ for the dust model consisting of carbonaceous grains and amorphous silicate grains from Draine \& Li (2007). The bump at $\lambda=0.2175 \mum$ arises from the absorption feature of carbonaceous grains and the bump at $\lambda=9.8 \mum$ is a silicate feature.}
\label{fig:f2}
\end{figure}

The optical depth per length along $\xhat$ and $\zhat$ directions are respectively given by
\bea
d\tau_{x}(\lambda)=\int_{a_{\min}}^{a_{\max}}\frac{dn}{da}C_{\ext}(\lambda,a)da
dx,\label{eq:dtaux}\\
d\tau_{z}(\lambda)=\int_{a_{\min}}^{a_{\max}}\frac{dn}{da}C_{\ext}(\lambda,a)da
dz,\label{eq:dtauz}
\ena 
where $C_{\ext}$ is the extinction cross section for a randomly oriented grain given in Equation (\ref{eq:Cext}) and $dn/da$ is the grain size distribution. For our modeling, we assume a dust model consisting of amorphous silicate grains and carbonaceous grains as in \cite{2007ApJ...657..810D} for $R_{V}=3.1$. If another grain size distribution is employed, it shall be clearly stated. The optical depth per gas column density for the adopted dust model is shown in Figure \ref{fig:f2}.

Provided the parameters of the illuminating star, one can solve Equation (\ref{eq:urad}) for $u_{\lambda}$ inside the cloud. The total energy density is obtained by integrating $u_{\lambda}$ over the entire spectrum of the stellar radiation field, i.e., $u_{\rad}(x,z)=\int u_{\lambda}d\lambda$.

\subsection{Critical size of aligned grains}

Using $u_{\rad}(x,z)$ for Equation (\ref{eq:Jmax_RAT}) one obtain $J_{\max}/J_{\th}$. The critical size $a_{\ali}(x,z)$ of aligned grains induced by RATs can be computed using the criteria in Section \ref{sec21}.

\subsection{Polarization Curves}\label{sec33}

Consider a column of dust along the $\zhat$ direction with the column density given by the visual extinction $A_{V}=1.086\tau_{z}(\lambda=0.55\mum)$. The alignment of grains along this line of sight is determined by $a_{\ali}(x,z)$.

The schematic illustration of calculations is shown in Figure \ref{fig:f1}. For simplicity, we assume that magnetic fields are uniform throughout the cloud and make an angle $\xi$ with the sky plane $\xhat\yhat$, while the starlight propagates along the $\zhat$ direction.

When $a_{\ali}(x,z)$ is known, the polarization arising from the aligned grains in a cell of $dz$ is computed as
\bea
dp_{\lambda}(x,z)=\int_{a_{\ali}}^{a_{\max}} \frac{\left(C_{x}-C_{y}\right)}{2}(dn/da)dadz,
\label{eq:dplam}
\ena
where the polarization arising from weakly aligned small grains $a<a_{\ali}$ is disregarded and graphite grains are assumed randomly oriented. It was shown in \cite{Hoang:2013tp} that small grains have small but finite residual alignment degree. However, they mostly affect the polarization of starlight in the UV wavelengths, whereas the peak of polarization is determined by aligned large grains.

Plugging Equation (\ref{eq:Cpol}) into this above equation, we obtain
\bea
dp_{\lambda}(x,z)=\int_{a_{\ali}}^{a_{\max}} C_{\pol}R(a)\cos^{2}\xi
\frac{dn}{da}dadz.
\ena

Using the Raleigh reduction factor $R$ from Equation (\ref{eq:Rayleigh}) for the above equation, we obtain
\bea
dp_{\lambda}^{\RAT}(x,z)=\int_{a_{\ali,\RAT}}^{a_{\max}} 
C_{\pol}\left[f_{\hi}^{\RAT}+(1-f_{\hi}^{\RAT})Q_{X,\lo}\right]\cos^{2}\xi\nonumber\\
\times \frac{dn}{da}dadz,~~~\label{eq:dpRAT}
\ena
where $Q_{X,\lo}$ is the degree of internal alignment with low$-J$ attractors (see Eq. \ref{eq:QXl}).

In general, $f_{\hi}$ is a function of grain size, grain shape, grain composition, and properties of the radiation field. To account for such a dependence, we introduce a parameter so-called the average fraction of grains aligned with high-$J$ attractor points, $\overline{f}_{\hi}$, and consider for different values of $\overline{f}_{\hi}$. 

Equation (\ref{eq:dpRAT}) is integrated over $z$ to obtain the polarization $p_{\lambda}(x)$ for each line of sight determined by the $x$ coordinate (or visual extinction $A_{V}$).

In the following, we apply our method to study grain alignment and predict polarization by aligned grains in the different environments that can be testable observationally. The description in this section is for a two dimensional cloud but it is easy to extend for three dimensional clouds (e.g., accretion disks and cometary coma).

\section{Grain alignment in Local Interstellar Medium}\label{sec:LISM}
\subsection{Local Interstellar medium and Local Radiation Field}

\begin{figure}
\includegraphics[width=0.45\textwidth]{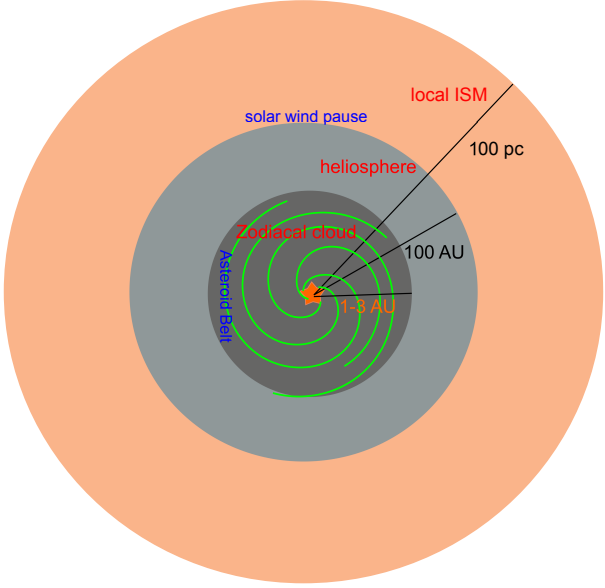}
\caption{Schematic illustration of the local interstellar medium, Zodiacal cloud and heliosphere. The Parker spiral magnetic field around the Sun is shown in green lines.}
\label{fig:LISM}
\end{figure}

LISM, extending from the Sun to the wall of Local Bubble at a distance $d\sim 100\pc$ (see Figure \ref{fig:LISM}), is an ideal location to test modern theories of grain alignment. Moreover, the linear polarization of nearby stars induced by aligned grains allows us to probe local magnetic fields and physical properties of the LISM (see e.g., \citealt{2012ApJ...760..106F}).

Observations in \cite{2012ApJ...760..106F} revealed a slow variation of the polarization with distance $d$ from the Sun for nearby stars within $d=5-40\pc$; although, their observations have strong fluctuations. On the other hand, measurements of polarization of nearby bright stars using a high precision polarimeter (PlanetPol) by \cite*{2010MNRAS.405.2570B} have showed that the polarization indeed increases with the increasing distance $d$, indicating the ubiquity of aligned grains in the LISM.
 
\subsection{Grain Alignment}
Let us discuss the possible alignment mechanisms in the LISM. Due to very low density, the LISM is optically thin, and the stellar radiation can freely propagate through the medium. Thus, we assume that the local ISRF can be described by the model of \cite*{1983A&A...128..212M}, which is the average radiation field in the solar neighborhood.

We assume physical parameters for the idealized LISM with the gas density $n_{\H}\approx 0.2 \cm^{-3}, T_{\gas}\approx 6000 \K$, although the LISM is known to be very patchy. The typical magnetic field $B=5\mu$G is expected (see a recent review by \citealt*{2011ARA&A..49..237F}). Using these typical parameters, we can estimate the timescale that the Davis-Greenstein mechanism brings grains to be aligned with the magnetic field:
\bea
 \tau_{\rm DG}\approx 1.2\times 10^{6} a_{-5}^{2} \yr,\label{eq:tauDG}
 \ena
whereas the gas bombardment tends to randomize the grain alignment over a timescale $\tau_{\gas}\approx 1.4\times 10^{6}\left(0.2\cm^{-3}/n_{\H}\right)a_{-5} \yr$. Thus, the Davis-Greenstein mechanism is inefficient for large grains with $a_{-5}>1$  (or $a<0.1\mum$) because $\tau_{\DG}>\tau_{\gas}$. Small grains of $a_{-5}<1$ are not efficiently aligned by the Davis-Greenstein mechanism either due to strong rotational damping by IR emission.

Indeed, using Equation (\ref{eq:FIR}), one can estimate the rotational damping coefficient by the infrared emission $F_{\rm IR}$ for the LISM condition as
\bea
F_{\rm IR}	= \frac{7.6}{a_{-5}}\left(\frac{u_{\rad}}{u_{\ISRF}}\right)^{2/3}
\left(\frac{0.2 \cm^{-3}}{n_{\H}}\right)~~~.\label{eq:FIRLISM}
\ena
For $a_{-5}<1$, the total damping rate for the grain becomes
\bea
\tau_{\rm drag}&=&\tau_{\gas}(1+F_{\rm IR})^{-1}\approx \tau_{\gas}F_{\rm IR}^{-1},\nonumber\\
&=& 1.1\times 10^{5}a_{-5}^{2}\left(\frac{u_{\rad}}{u_{\ISRF}}\right)^{-2/3} \yr. \label{eq:taudrag}
\ena

Comparing $\tau_{\rm drag}$ with $\tau_{\rm DG}$ from Equation (\ref{eq:tauDG}), it turns out that grains cannot be aligned by the Davis-Greenstein mechanism since $\tau_{\rm drag}\ll \tau_{\DG}$.

Using Langevin equations, Hoang et al. (2014) studied the alignment of small grains ($a<0.1\mum$) by the paramagnetic relaxation, taking into account the various damping and excitation processes. They found that the degree of paramagnetic alignment of such small grains is at most of 5$\%$ for the conditions of warm ionized medium (e.g., similar to LISM) with the magnetic field $B=10\mu$G. Obviously, the RAT alignment appears to be the principal mechanism that drives grain alignment in the LISM.

From Equations (\ref{eq:Jmax_RAT}) and (\ref{eq:taudrag}), one can estimate the maximum rotation rate induced by RATs for the cases in which the grain rotational damping is dominated by gas collisions and IR emission. These are respectively given by
\bea
\frac{ J_{\max}^{\RAT}}{J_{\th}}&\approx& 500\hat{\gamma}_{\rad}a_{-5}^{1/2}\left(\frac{\overline{Q}_{\Gamma}}{10^{-2}}\right),
{\rm for~} a >10^{-5}\cm,~~~\\
\frac{J_{\max}^{\RAT}}{J_{\th}} &\approx& 66\hat{\gamma}_{\rad} a_{-5}^{3/2}\left(\frac{\overline{Q}_{\Gamma}}{10^{-2}}\right),
{\rm for~} a <10^{-5}\cm.~~~\label{eq:JmaxIR}
\ena
It can be seen that small grains of $a_{-5}\sim 1$ can still rotate suprathermally thanks to RATs.

\subsection{Polarization by aligned grains due to RATs}

To calculate the polarization of starlight by aligned grains in the LISM, we model it as a thin disk of the heliocentric radius $R_{\rm helio}=100$ pc. It is worth noting that the magnetic field direction in the LISM has been studied through the polarization of nearby stars and is found to be close to the direction of the interstellar magnetic field (i.e., parallel to the Galactic plane; \citealt{2011ARA&A..49..237F}). However, detailed information on the magnetic field along the entire line of sight to a particular star is unknown. Thus, we consider the magnetic field in the galactic plane but incorporate the magnetic geometry along the line of sight into the effective degree of alignment, $R_{\rm eff}=R\cos^{2}\xi$.

We take the local radiation field from \cite{Mezger:1982p4014} with $u_{\rad}=u_{\rm IRSF}=8.64\times 10^{-13}\erg \cm^{-3}$. From Equation (\ref{eq:JmaxIR}), one can derive the critical size of aligned grains $a_{\ali}\approx 0.07\mum$ for the LISM, which does not depend on the gas density $n_{\gas}$ due to the dominance of damping by IR emission. Note that $\overline{Q}_{\Gamma}$ decreases rapidly with $a$. For the ISRF, we estimate $\overline{Q}_{\Gamma} \approx 2.4\times 10^{-3} \left(\bar{\lambda}/1.2\mum\right)^{-2.7}a_{-5}^{2.7}$ for $a \ll \overline{\lambda}$.

\begin{figure}
\includegraphics[width=0.45\textwidth]{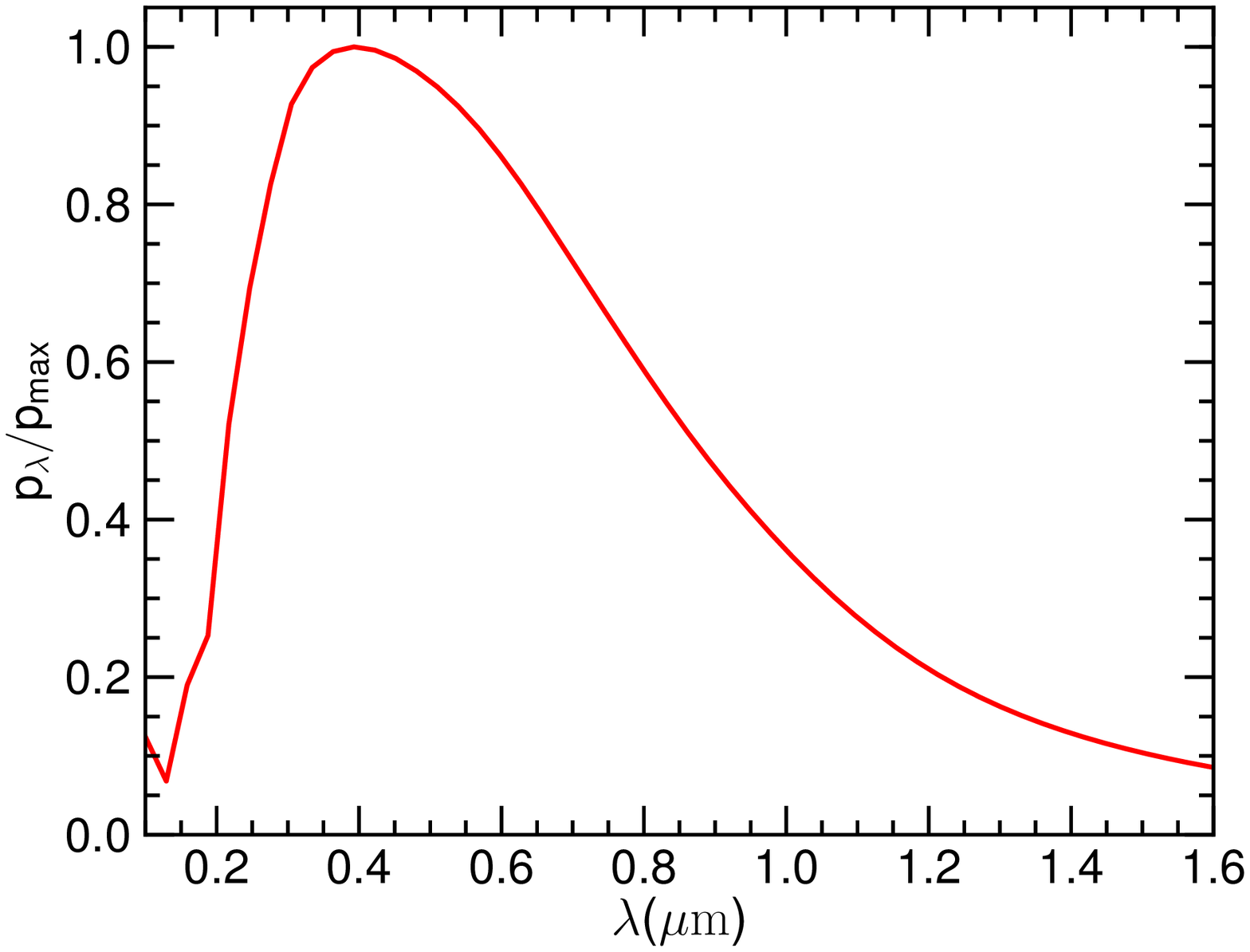}
\includegraphics[width=0.45\textwidth]{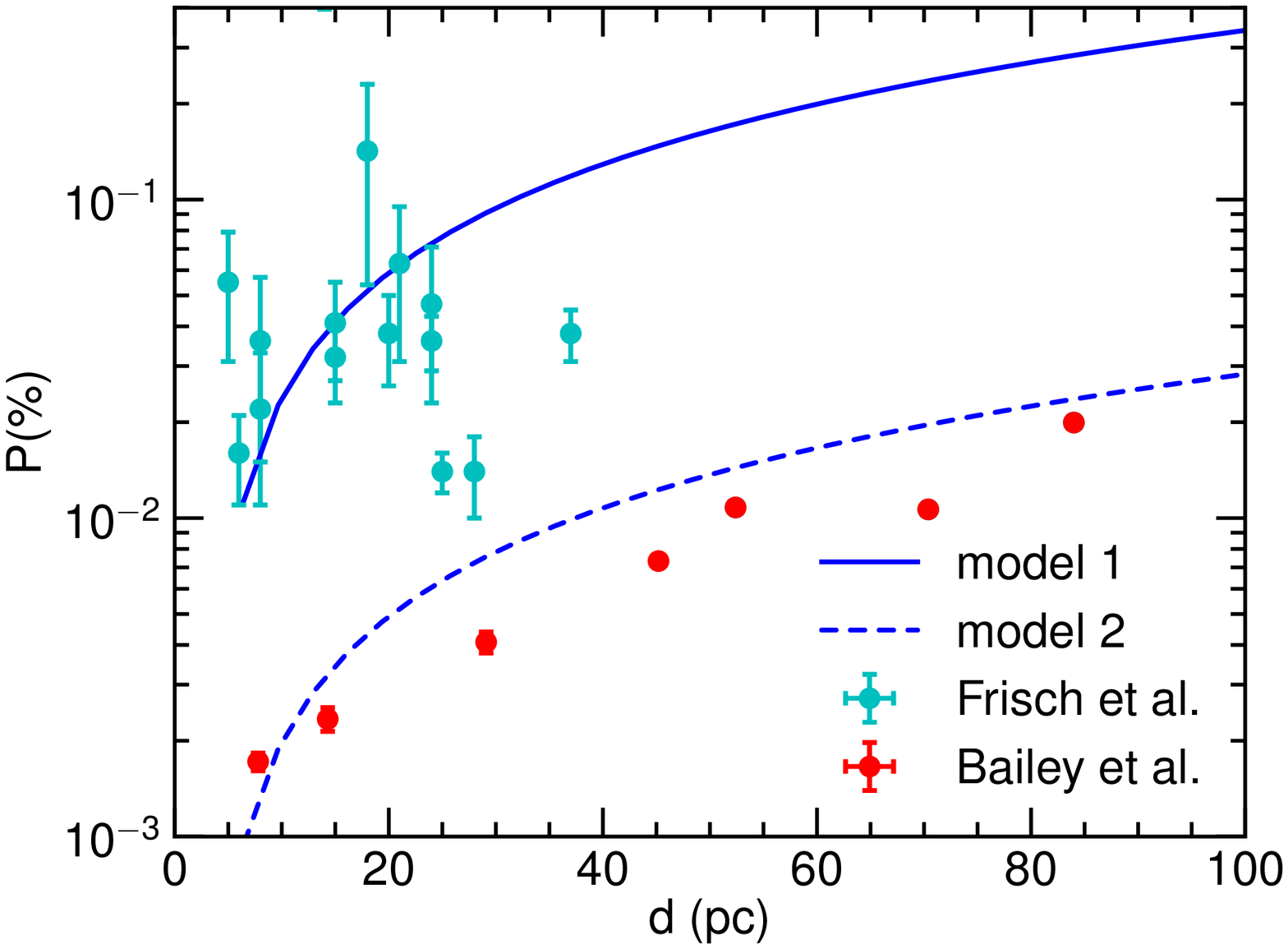}
\caption{{Upper:} polarization curve of starlight induced by aligned grains in the LISM. {Lower:} maximum polarization of nearby stars as a function of distance from the Sun from our predictions (see the text) for model 1 (solid line) and model 2 (dashed lines). The observational data for the stars with R. A. $<$ 17 h from Bailey et al. (2010) are shown in red filled circles, and data from Frisch et al. (2012) are shown in cyan filled circles. The predicted polarization increases with the increasing distance $d$ of the stars.}
\label{fig:LIC}
\end{figure}

Figure \ref{fig:LIC} (upper) shows the polarization curve arising from grains aligned in the local magnetic field due to RATs. The wavelength at which the polarization peaks, $\lambda_{\max}$, is equal to $0.4\mum$.

Figure \ref{fig:LIC} (lower) shows the maximum polarization predicted for two models with $n_{\gas}=0.2 \cm^{-3}$ (solid line, model 1) and $n_{\gas}=0.015~\cm^{-3}$ (dashed line, model 2) for $R\cos^{2}\xi=0.9$. The polarization data of the stars with R. A. $<$ 17 h from \cite{2010MNRAS.405.2570B} are shown in filled red circles and the data from \cite{2012ApJ...760..106F} are shown in cyan circles. Our models predict the increase of polarization with the increasing distance to the stars. It can be seen that model 2 is in a good agreement with the PlanetPol data. The observational data by \cite{2012ApJ...760..106F} exhibit large scatter, so it is difficult to conclude that their data are consistent with our simple modeling. It is noted that the level of polarization is proportional to the dust column density of the line of sight. Thus, the polarization for stars with RA $<$ 17 h in \cite{2010MNRAS.405.2570B} is much lower than that for the stars in \cite{2012ApJ...760..106F} because the regions with RA $<$17 h are expected to have the lower dust content. 

\section{Grain alignment in accretion disks}\label{sec:accdisk}
Magnetic fields are widely believed to play an important role in accretion disks, but understanding to what extent polarimetry traces magnetic fields in dense regions is still very limited. \cite{2007ApJ...669.1085C} (hereafter CL07) modeled polarized emission from aligned grains in a T Tauri disk and predicted a polarization level of $2\%-3\%$ at the wavelength $\lambda=100\mum$. Recently, the Submillimeter Array (SMA) observations (\citealt{2009ApJ...704.1204H}; \citealt{2013AJ....145..115H}) show no considerable level of polarization from some T Tauri disks. In this section, we revisit the problem of grain alignment in the accretion disks by taking into account the effects that were neglected in the CL07 model, including the dependence of grain alignment on the radiation anisotropy direction and with the magnetic field and the assumption of perfect alignment of grains, and predict an upper limit of polarization of thermal emission by aligned grains.

\subsection{Disk Model Assumptions}
\begin{figure}
\includegraphics[width=0.48\textwidth]{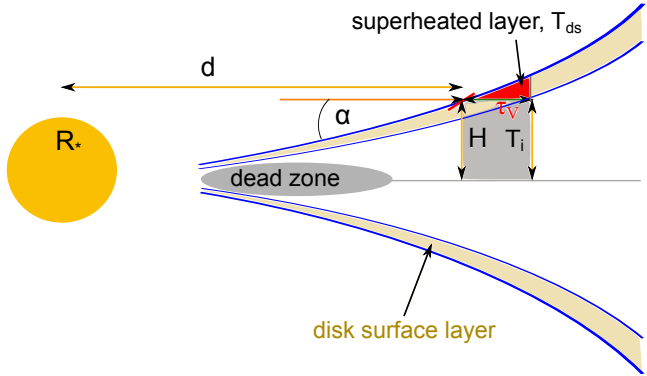}
\caption{Schematic illustration of a flared disk around protostar consisting of dead zone, surfaces layers and disk interior. Surface layers are directly heated by stellar radiation, which is able to penetrate a thickness of optical depth $\tau_{V}=1$, to temperature $T_{ds}$, and the re-emission of the hot dust grains in surface layers will heat the disk interior to $T_{i}$.}
\label{fig:flaredisk}
\end{figure}

We consider the model of a typical accretion disk consisting of a central star surrounded by a flared disk (see Fig. \ref{fig:flaredisk}). A detailed description of the disk model is presented in \cite{1997ApJ...490..368C}, here we summarize the essential parts. At a distance $d$ from the star, the surface layers are heated to a temperature $T_{ds}$ by radiation from the central star. Dust grains in these superheated layers reemit radiation in IR, which is transparent throughout the disk, and in turn IR emission heats gas and dust in the disk interior to a temperature $T_{i}$. Essentially, the radiation field in the disk interior comprises attenuated radiation from the central star and superheated layers (of temperature $T_{ds}$), and the IR emission from dust (of temperature $T_{i}$). As a result, the spectral energy density (SED) of T-Tauri disks usually has a bump at near-IR, corresponding to the peak of stellar blackbody emission, and a tail in far-IR due to the re-emission of heated dust from the disk. Temperature profiles $T_{ds}$ and $T_{i}$, height of the disk $H$, and other properties of the disk model are taken from \cite{1997ApJ...490..368C}.

We assume that the gas distribution decreases exponentially from the disk plane, governed by
\bea
n_{\gas}(x,y,z)=n_{\gas}(x,y,z=0)\exp\left(-\frac{z^{2}}{2h^{2}}\right),\label{eq:ngas}
\ena
where $h$ is the gas scale height from the mid-plane ($z=0$).

The mass column density of the disk is taken as $\Sigma=\Sigma_{0}\left(d/{\rm AU}\right)^{-3/2}$, where $d$ is the disk radius and $\Sigma_{0}=1000\g\cm^{-2}$. From $\Sigma$, one can derive the gas density distribution within the disk, $n_{\gas}$, as follows:
\bea
\Sigma&=&\int m_{\H}n_{\gas}(\rho,z)dz=n_{\gas}(\rho,z=0)2\int_{0}^{h} \exp-\left(\frac{z}{\sqrt{2}h}\right)^{2}dz,\nonumber\\
&=&2^{-1/2}n_{\gas}(\rho,z=0)H\erf(2^{-1/2}),
\ena
where $h=H/4$ has been used. Thus,
\bea
n_{\gas}(x,y,z=0)&\approx&\frac{1}{0.5m_{\H}}\frac{\Sigma}{H}=\frac{\Sigma_{0}}{m_{\H}}\left(\frac{d}{{\rm AU}}\right)^{-3/2},\\
&\approx&4.0\times 10^{14}\left(\frac{d}{\rm AU}\right)^{-39/14} \cm^{-3}. 
\ena

\subsection{Critical size of aligned grains}
Grain size distribution is assumed to follow the power law (\citealt{Mathis:1977p3072}) having a lower cutoff $a_{\min}=0.05\mum$ and an upper cutoff $a_{\max}$. Although very big grains are likely present in the accretion disk due to coagulation, conservatively, we assume $a_{\max}=1\mum$ as in \cite{1997ApJ...490..368C}.

Unlike CL07, who approximated the stellar radiation field as a monochromatic field with $\lambda=\lambda_{\max}$ and computed $J^{\RAT}/J_{\th}$ for this wavelength, here, we integrate over the entire spectrum of the radiation field $u_{\lambda}$ for each position within the disk.

\begin{figure}
\includegraphics[width=0.45\textwidth]{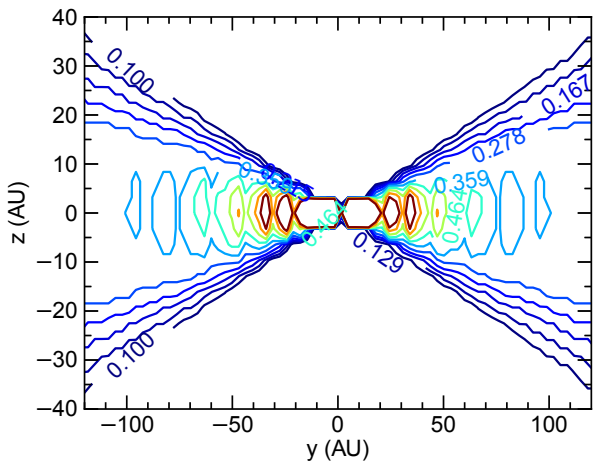}
\caption{Contours of critical size of aligned grains, $a_{\ali}$, in the plane $yz$ with $x=0$ of the accretion disk. Small grains ($a\sim 0.1\mum$) in the surface layers can be aligned by the stellar radiation. In the disk interior, only large grains ($a\sim 1\mum$) can be aligned.}
\label{fig:AD}
\end{figure}

Figure \ref{fig:AD} shows the contours of $a_{\ali}$ in the plane of $yz$ perpendicular to the disk midplane. It can be seen that grains with $a\ge a_{\ali}\sim 0.1\mum$ can be aligned in the surface layers. In the disk interior, $a_{\ali}$ increases, indicating that only large grains can be aligned. 

\subsection{Effects of radiation field and magnetic fields}
In the model of a flared disk, the surface layers are directly illuminated by stellar radiations. The thickness of this layer corresponds to $\tau_{V}=1$. Hot dust grains that are heated in the surface layers will heat dust and gas in the disk interior. 

Due to large optical depth, dust grains in a given cell of the disk interior receive photons mostly from the heated surface layers arriving from the vertical direction perpendicular to the disk plane. If the magnetic field is toroidal (azimuthal) parallel to the disk plane as suggested by observations (\citealt{Tamura:1999p6500}), then the radiation direction is perpendicular to the magnetic field (i.e., $\psi=90^{\circ}$) both in the surface layer and disk interior. Therefore, we expect that grains are weakly aligned.

For the alignment with $\psi=90^{\circ}$, grains are aligned with $J\sim J_{\rm th}$ regardless of radiation intensity and grain size (see \citealt{2008MNRAS.388..117H}). As a result, the fraction of grains aligned with high-$J$ attractor points $f_{\hi}\approx 0$.

In realistic conditions, the magnetic field is not perfectly azimuthal due to disk instability. In addition, the dust grains in the disk interior receive the radiation from the surface layers coming from a wide range of angle with the intensity decreasing with the increasing angle between the radiation direction and the vertical axis. Therefore, grains can be aligned with the magnetic field, but the alignment efficiency would be much lower than for the ideal situation in which the radiation is parallel to the magnetic field. A detailed study using simulated magnetic field data is beyond the scope of this paper.

\subsection{Maximum polarization of thermal dust emission}
To estimate the maximum polarization of thermal dust emission, we disregard the dependence of magnetic field geometry and anisotropic direction of radiation. Thus, the polarization can be given by
\bea
P_{\rm em}\equiv \frac{I_{\lambda,\pol}}{I_{\lambda}}=\frac{\int _{a_{\ali}}^{a_{\max}} C_{\pol}R(a)n(a) da}{\int _{a_{\min}}^{a_{\max}} C_{\ext}n(a) da},\label{eq:pmax}
\ena
where $I_{\lambda,\pol}$ is the polarized emission, $I_{\lambda}$ is the total dust emission, $C{\ext}$ and $C_{\pol}$ are the extinction and polarization cross section, and $R(a)$ is given by Equation (\ref{eq:Rayleigh}).

\begin{figure}
\includegraphics[width=0.45\textwidth]{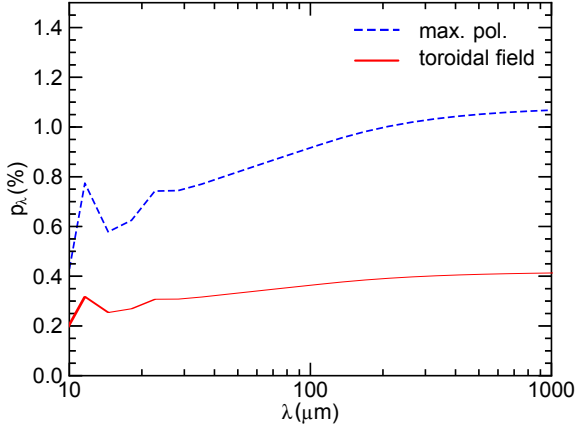}
\caption{Maximum degree of polarization for thermal emission from the disk (dashed line) and the expected polarization when the radiation anisotropy direction perpendicular to the magnetic field (solid line). The degree of polarization is much lower when the angle dependence of RAT alignment is accounted for.}
\label{fig:polem}
\end{figure}

Figure \ref{fig:polem} (dashed line) shows the maximum polarization degree of thermal emission from the disk assuming that the magnetic field is always parallel to the direction of radiation source. We assume that all grains larger than $a_{\ali}$ are perfectly aligned (i.e., $f_{\hi}=1, R(a)=1$). As shown, the degree of polarization is essentially less than $1.1\%$ and rises slowly with the increasing $\lambda$ for $\lambda>20\mum$. The solid line in Figure \ref{fig:polem} shows the expected polarization for the case of the toroidal magnetic field (i.e., $\psi=90^{\circ}$ and $f_{\hi}=0, R(a)=Q_{X,\lo}$). It can be seen that the maximum polarization is decreased substantially.

\subsection{Alignment of large grains and internal relaxation}
Large grains with $a>1\mum$ are expected to have negligible internal relaxation. The RAT alignment of large grains is determined by the helicity of grain shape, therefore, the RAT alignment of large grains is shown to be similar to the alignment of interstellar grains (see \citealt{2009ApJ...697.1316H}), in which grains can be aligned with both high-$J$ and low-$J$ attractor points. The only difference is that for large grains without internal relaxation, the alignment with low-$J$ attractor points can occur with grain long axes parallel to the angular momentum (also magnetic field). This new situation makes the modeling of polarized emission by large grains in accretion disks more complicated.

\section{Grain alignment in interplanetary medium and Circular polarization}\label{sec:cirpol}

\subsection{Theoretical Basis}

CP of starlight in general can be produced by (i) multiple scattering of unpolarized light by anisotropic medium (\citealt*{1972ApJ...177..177K}), (ii) single scattering by aligned grains (see \citealt{1973MNRAS.162..367B}; \citealt{1973IAUS...52..131S}; \citealt{1978A&A....69..421D}), and (iii) scattering by optically active chiral particles (see e.g., \citealt*{2004IAUS..213..149W}). \cite{1972MNRAS.159..179M} proposed an elegant mechanism of producing circular polarization based on the variation of grain alignment direction along a line of sight. The idea was later extended in \cite{Martin:1976p7959}. The idea is that, when the starlight passes through a layer of aligned dust, it becomes partially linearly polarized. By passing through the second layer in which the alignment direction of grains makes some angle with that in the first layer, the light becomes circularly polarized. If on passing through a single layer, the linear polarization degree is equal to $p$, then passing through two layers with the different alignment directions produces CP that does not exceed $p^2$.

Literature study shows that the multiple scattering mechanism, which requires optically thick (i.e., optical depth $\tau>1$) environments, is well remembered (see \citealt{1988ApJ...326..334B}), while the process of single scattering by aligned grains is frequently forgotten. However, for optically thin environments (i.e., $\tau<1$), e.g., ZC and comets, the double and multiple scattering are rare. Thus, the single scattering by aligned grains appears to be the most promising mechanism responsible for the observed CP.

Assuming that the incident light is unpolarized with intensity $I_{0}$, the intensity of light scattered by a randomly oriented grain towards an observer at distance $r$ is given by
\bea
I_{\rm sca}=\frac{I_{\rm 0}}{k^{2}r^{2}}\langle S_{11}\rangle=\frac{I_{\rm 0}}{r^{2}}\sigma_{\rm sca},
\ena
where $k=2\pi/\lambda$, $\langle S_{11}\rangle$ is the oriental averaging of the first element of the Muller matrix, and $\sigma_{\sca}=Q_{\sca}/\pi a^{2}$ is the scattering cross section (see \citealt{1983asls.book.....B}).

The intensity of radiation scattered from a volume $\Delta \Gamma$ of dust grains at distance $R$ from the Sun and $r$ from the observer is equal to
\bea
\Delta I ({\Rv}, {\rv})=\frac{L_{\star} n_{\d}\sigma_{\sca}}{4\pi R^2
r^2}\Delta \Gamma~~~,
\label{eq:dI}
\ena
where $I_{0}=L_{\star}/4\pi R^{2}$ with $L_{\star}$ being the stellar luminosity, $n_{\d}$ is the number density of dust grains of a given size. Above, the vector $\Rv$ is directed from the Sun and $\rv$ is directed from the observer, and $|\Rv-\rv|=1\AU$.

The intensity of circularly polarized radiation due to single scattering by an aligned grain is given by
\bea
V(\be_{0},\be_{1},\be)=\frac{1}{k^{2}r^{2}}S_{41}I_{0},
\ena
where $S_{41}$ is one element of the Muller scattering matrix (\citealt{1983asls.book.....B}) .

For grains small compared to the wavelength (Rayleigh limit), i.e., $ka=2\pi a/\lambda\ll 1$, the Stokes $V$ parameter has been derived by several authors using the dielectric dipole approximation (\citealt{1973MNRAS.162..367B}; \citealt{1973IAUS...52..131S}; \citealt{1978A&A....69..421D}; \citealt{2000MNRAS.314..123G}). Thereby,
\bea
V=\frac{I_{0}k^{4}}{2r^{2}}i\left(\alpha_{\|}\alpha_{\perp}^{*}-\alpha_{\|}^{*}\alpha_{\perp}\right).\left([\be_{0}\times\be_{1}].\be\right)\left(\be_{0}.\be\right),~~~\label{eq:V}
\ena
where $\be_{0}={\bk}_{0}/k, \be_{1}={\bk}_{1}/k$ are the unit vectors of incident and scattering direction, $\alpha_{\|}$ and $\alpha_{\perp}$ are the complex polarizabilities along the grain symmetry axis $\be$ and in the perpendicular direction, respectively (see Appendix \ref{apdx:b}). 

The intensity of circularly polarized radiation scattered by a volume $\Delta \Gamma$ of dust at distances $R$ from the star and $r$ to the observer reads
\bea
\Delta V ({\Rv}, {\rv})=\frac{L_{\star} n_{\d}\sigma_{V}}{4\pi R^2
r^2}A \left(\left[\frac{\Rv}{R}\times \frac{\rv}{r}\right].{\bf h}\right)
\left(\frac{\Rv}{R}.{\bf h}\right)\Delta \Gamma~~~,
\label{eq:dV}
\ena
where $A=\langle(\be.{\hv})^{2}\rangle-1/3$ is a parameter describing the alignment of grain symmetry axis $\be$ with the axis of alignment $\hv$ and
\bea
\sigma_V=\frac{1}{2}ik^4\left(\alpha_{\|}\alpha^{\ast}_{\bot}-\alpha^{\ast}_{\|}\alpha_{\bot}\right),
\label{eq:sigmaV}
\ena
is the CP cross section.

From the above equation one can see that CP is only produced when the grain is an absorbing material, i.e, its polarizability (or refractive index) contains a non-zero imaginary part. Moreover, one can see that CP is not equal to zero only when there exists a magnetic field component ${\bf h}_{\perp}$ perpendicular to the ecliptic plane as determined by $\Rv\times \rv$. Thus, the non-zero observed CP of scattered light reveals that the perpendicular component is indeed present.

To calculate $\sigma_{V}$, we assume the dielectric function $m=1.7-0.1i$ for silicate grains. For small grains $ka\ll 1$, Equation (\ref{eq:sigmaV}) is employed to calculate $\sigma_{V}$ with $\alpha_{\|}$ and $\alpha_{\perp}$ obtained from Equation (\ref{eq:alpha}). Otherwise, the approximation of $\sigma_{V}$ in \cite{1978A&A....69..421D} is adopted for simplicity.

\begin{figure}
\includegraphics[width=0.45\textwidth]{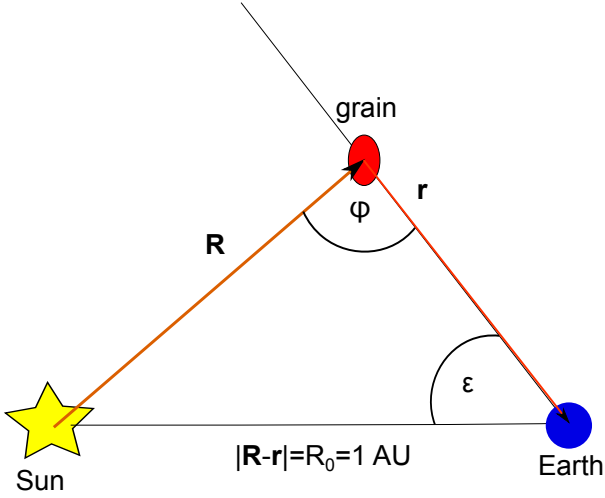}
\caption{Geometry of calculations. A dust grain aligned with the magnetic field scatters the sunlight towards the Earth.}
\label{fig:sketch}
\end{figure}

\subsection{Dynamics and lifetime of dust in interplanetary medium}
Dust in interplanetary medium is subject to gravity of the Sun, radiative pressure by sunlight, molecular force by solar winds, and Poynting-Robertson (PR) drag. The radiative force is equal to
\bea
F_{\rm rad}=\frac{L_{\odot}}{4\pi d^{2}c}\langle Q_{\rm pr}(a,\lambda)\rangle \pi a^{2},\label{eq:Frad}
\ena
where $Q_{\rm pr}(a,\lambda)=Q_{\abs}+Q_{\sca}(1-\cos\theta)$ is the radiative pressure cross-section and $\langle Q_{\rm pr}\rangle=\int Q_{\rm pr}u_{\lambda}d\lambda/u_{\rad}$ denotes the averaging over the spectrum of the solar radiation field.

The ratio of radiative force to gravity force is given by
\bea
\beta&=&-\frac{F_{\rad}}{F_{\rm gra}}=\frac{3L_{\odot}\langle Q_{\rm pr}(a,\lambda)\rangle}{16\pi G M_{\odot}c}(\rho a)^{-1}\\
&\approx &1.91\left(\frac{\langle Q_{\rm pr}(a,\lambda)\rangle}{1.0}\right)\hat{\rho}^{-1} a_{-5}^{-1}.\label{eq:beta}
\ena

Equation (\ref{eq:beta}) reveals that $\beta$ is determined by the grain size and $\langle Q_{\rm pr}\rangle$, regardless of its distance to the Sun. For the mean wavelength of the sunlight $\bar{\lambda}=0.6\mum$, $\langle Q_{\rm pr}\rangle$ is in the range $1-2$ for $0.1\mum$ grains. Therefore, the $0.1\mum$ grains are easily expelled from the ZC by the radiative pressure. The value of $\beta$ is smaller than 1 for big ($\sim 1\mum$) grains due to the increase of $a$ (increased gravity) while $Q_{\rm pr}$ is mostly saturated for $\lambda<<a$. Very small grains also have $\beta< 1$ because $Q_{\rm pr}$ decreases slowly (see \citealt{Burns:1979bg}; \citealt{1994AREPS..22..553G}). As a result, both very small and big grains are difficult to be blown away by the sunlight.

The collisions of ions in the solar winds with the grain also result in drag. The drag force due to both neutral and ions is given by
\bea
F_{\rm sw}=n_{\gas}k_{\B}T_{\gas} \pi a^{2} C,\label{eq:Fsw}
\ena
where $n_{\gas}$ is the gas number density and $C$ is the drag coefficient (\citealt{Baines:1965p3201}; \citealt{1979ApJ...231..438D}).

The ratio of the drag force to gravity is equal to
\bea
\beta_{\rm sw}&=&\frac{F_{\rm sw}}{F_{\rm gra}}\nonumber\\
&\approx& 3.5\times 10^{-5}\left(\frac{n_{\gas}}{10^{2}\cm^{-3}}\right)\left(\frac{T_{\gas}}{100\K}\right)\left(\frac{C}{10}\right)\left(\frac{r}{\AU}\right)^{2}a_{-5}^{-1}.~~~~
\ena

The above equation shows that the force due to solar winds is much smaller than the radiative force for the diffuse clouds.

Micron-sized dust grains are mostly affected by the PR drag force. \cite{Burns:1979bg} showed that the PR force acts to gradually decrease the semimajor axis and eccentricity of the orbit of micron-sized grains around the Sun, resulting in the loss of micron-sized grains. The orbital decay time for a dust grain on circular orbit is given by
\bea
\tau_{\rm PR}=\frac{R^{2}}{4\eta \langle Q_{\rm pr}\rangle}\approx 210 a_{-5} \hat{\rho} \frac{R_{\AU}^{2}}{\langle Q_{\rm pr}\rangle} \yr,\label{eq:tau_PR}
\ena
where $\eta=S_{0}r_{0}^{2}A/mc^{2}=2.53\times 10^{11}(\rho a)^{-1}$ with $S_{0}$ being the solar flux at distance $r_{0}$, $R_{\AU}$ is the semimajor axis of the grain elliptical orbit in units of AU.

\subsection{Zodiacal cloud}\label{sec7}
\subsubsection{Introduction}

ZC is the interplanetary medium extending from the Sun to Jupiter. It can be seen by naked eyes as a glow triangle originating from the scattering of sunlight by dust grains just before the sunrise and after the sunset. Zodiacal dust is believed to be produced by the evaporation of comets when approaching the Sun and by collisions of comets with the asteroid belt. 

Thermal emission from Zodiacal dust is an important foreground component that contaminates to the CMB signal and has been studied extensively (\citealt{1998ApJ...508...44K}; \citealt{Collaboration:2013uoa}). However, whether Zodiacal emission is polarized and what is the degree of polarization remain unclear. In principle, we expect to see polarized Zodiacal emission if dust grains can be aligned.

As discussed in the previous section, a conclusive evidence of aligned grains can be sought for through the CP of Zodiacal light. Observationally, CP of Zodiacal light has been reported by several groups (\citealt{1972ApJ...176L.115K}; \citealt{Staude:1972ud}; \citealt{1972ApJ...177L.137W}). 

A special feature in CP of Zodiacal light is that it changes sign at the elongation angle $\epsilon \approx 80^{\circ},~ 180^{\circ}$ and $280^{\circ}$. \cite{1976Ap&SS..43..257D} suggested that grains can be aligned with the IMF due to solar winds or radiative flux and explained such a feature by means of a compound model of grain alignment. According to \cite{1976Ap&SS..43..257D}, Zodiacal dust comprises both prolate and oblate spheroidal grains, and their different alignment with the magnetic field could explain the observed CP. However, modern RAT alignment theory indicates that the alignment of prolate and oblate grains is similar in the sense that their axis of maximum moment of inertia tends to align with the magnetic field. Below, we consider oblate grains and improve the Dolginov \& Mytrophanov treatment. 

\subsubsection{Physical Model}
We consider a simplified physical model of ZC in which ZC is modeled as a thin disk with heliocentric radius $R=3\AU$. The gas density decreases with the increasing distance from the Sun as a power law:
\bea
n_{\gas}(r)=r^{-\eta},
\label{eq:K}
\ena
where $\eta=1.34$ is assumed as in \cite{1998ApJ...508...44K}.

As in \cite{1978A&A....69..421D}, we assume that the size distribution of Zodiacal dust follows a power law, $dn/da\propto a^{\alpha}$ with the typical value $\alpha=-3.5$ as in the ISM (\citealt*{Mathis:1977p3072}). Based on fitting to the observational data of COBE, \cite{2002ApJ...578.1009F} found that the size distribution of silicate has a breakup at $a\sim 32\mum$. We note that small dust grains in ZC can be produced due to collisions of comets with asteroids, and have been observed (see e.g., \citealt{LeChat:2013ut}).

The magnetic field in the ZC is assumed to be the Archimedean spiral, as proposed by \cite{1958ApJ...128..664P} (Parker model). The spiral magnetic field is believed to originate from magnetic fields that are frozen into plasma and carried away by solar winds. It is noted that the classic model of the IMF by \cite{1958ApJ...128..664P} does not have a vertical component $B_{\perp}$, but the existence of a rapidly fluctuating component $B_{\perp}$ has been reported by the Ulysses spacecraft (\citealt{1996JGR...101..395F}). This fast changing component $B_{\perp}$ is thought to arise from waves and disturbances of solar winds, and depends on the variation of the horizontal component. Thus, we assume that the magnetic field in the interplanetary medium consists of a spiral component in the ecliptic plane $B_{\|}$ and a component perpendicular to it, $B_{\perp}$. 

Let $\theta$ be the angle between the magnetic field and the ecliptic plane. The magnetic field at a heliodistance $r>r_{0}$ consists of the radial, azimuthal and vertical components:
\bea
B_{r}(r,\phi,\theta)&=&B_{0}\left(\frac{r}{r_{0}}\right)^{-2},\\
B_{\phi}(r,\phi,\theta)&=&B_{0}\left(\frac{r}{r_{0}}\right)^{-1}\cos\theta,\\
B_{\theta}(r,\phi,\theta)&=&B_{0}\sin\theta.\label{eq:BParker}
\ena
where $B_{0}$ is the magnetic field at $r=r_{0}$. The first two components come directly from the Parker model, and the last one represents a potential vertical component. $B_{0}\approx 30\mu$G
at $r_{0}=1\AU$, although this value may vary with the solar activity. 

Physical parameters and characteristic timescales of grain dynamics in ZC are given in Table \ref{tab:ZC}.

\subsubsection{Alignment of Zodiacal dust}

The alignment of Zodiacal dust is complicated because it involves various mechanisms, including the Davis-Greenstein mechanism, RAT alignment by solar radiation, and mechanical alignment by solar winds. In the following, we discuss the importance of these alignment mechanisms.

First of all, one can realize that the ZC is rather dilute and hot, with the gas density $n_{\H}\sim 5\cm^{-3}$, temperature $T_{\gas}=10^{4}$ K (see \citealt{1978A&A....69..421D}). The energy density of solar radiation at the location of ZC is $u_{\rad}\approx 2\times 10^{-4} \erg\cm^{-3}$. With these parameters, one can estimate the rotational damping time due to the gas bombardment and IR emission:
\bea
\tau_{\rm drag}&=& \tau_{\gas}\left(1+F_{\rm IR}\right)^{-1},\\
&\approx&\tau_{\gas}F_{\rm IR}^{-1}
\approx8.0\times 10^{4}a_{-5}^{2}\left(\frac{u_{\rad}}{u_{\ISRF}}\right)^{-2/3},\nonumber\\
&\approx &0.21a_{-5}^{2}\yr,\label{eq:tau_damp}
\ena
where $F_{\rm IR}$ from Equation (\ref{eq:FIR}) has been used.

With the magnetic field $B \approx 50\mu$G in ZC, the Davis-Greenstein alignment timescale is equal to
\bea
 \tau_{\DG}\approx 3.2\times 10^{3}a_{-5}^{2}\yr,
 \ena
for normal paramagnetic grains. Because $\tau_{\DG}\gg \tau_{\rm drag}$, the Davis-Greenstein mechanism is inefficient for aligning paramagnetic grains in the ZC.

Secondly, dust grains in ZC expose to solar winds of high velocity $v_{\rm flow}=300-400\km\s^{-1}$. The supersonic flows are expected to drive grains to be aligned with their shortest axes perpendicular to the flow according to the Gold mechanical mechanism. The timescale of the Gold alignment can be estimated to be equal to the spin-up timescale $\tau_{\rm Gold}\approx 4a_{-5}\yr$ (see Eq. \ref{eq:tauGold}). Comparing $\tau_{\rm Gold}$ with $\tau_{\rm drag}$, one can clearly see that $\tau_{\rm Gold}> \tau_{\rm drag}$. Therefore, the Gold alignment mechanism is inefficient in aligning Zodiacal dust grains.

Thirdly, solar radiation at a distance of $1\AU$ has the energy density $u_{\rad}$ and mean wavelength $\bar{\lambda}\approx 0.9 \mum$. From Equation (\ref{eq:Jmax_RAT}) with the use of Equation (\ref{eq:tau_damp}), one can estimate the maximum rotational rate of grains as
\bea
\frac{J_{\max}^{\RAT}}{J_{\th}}&\approx& 200\hat{\gamma}_{\rad}a_{-5}^{1/2}\left(\frac{u_{\rad}}{u_{\ISRF}}\right)F_{\rm IR}^{-1},\nonumber\\
&\approx& 1.4\times 10^{3}\hat{\gamma}_{\rad}a_{-5}^{3/2}\left(\frac{\overline{Q}_{\Gamma}}{10^{-3}}\right).
\ena

Given strong solar radiation field ($u_{\rad}\gg u_{\rm ISRF}$), small grains of $0.01\mum$ can still be spun-up to suprathermal rotation by RATs (see Eq. \ref{eq:Jmax_RAT}). Indeed, the timescale for RATs to spin-up grain to suprathermal rotation is equal to
\bea
 \tau_{\RAT}&=&\frac{\tau_{\rm drag}}{J_{\max}^{\RAT}/J_{\th}}=7.4\times 10^{-4}\hat{\gamma}_{\rad}^{-1}a_{-5}^{-3/2}\left(\frac{\overline{Q}_{\Gamma}}{10^{-3}}\right)^{-1}\tau_{\drag}\nonumber\\
 &\approx& 1.4\times 10^{-4} \hat{\gamma}_{\rad}^{-1}\left(\frac{\overline{Q}_{\Gamma}}{10^{-3}}\right)^{-1}a_{-5}^{1/2}\yr.\label{eq:taurat}
\ena

It is clearly seen that $\tau_{\RAT}<\tau_{\rm Gold}$, i.e., the spin-up by RATs dominates that by gaseous flows for $a_{-5}>1$ grains. It is worth noting that the timescales of alignment are much shorter than the orbital decay time due to PR drag force (Equation \ref{eq:tau_PR}), indicating that grains are well aligned before they are captured by the Sun.

Finally, when the grain rotation is spun-up by RATs, the direction of grain alignment, whether with respect to the radiation direction or magnetic fields, depends on the precession timescales of grain angular momentum around these axes. For normal paramagnetic material, the Larmor precession time of the angular momentum around the magnetic field $\tau_{B}$ is long compared to the precession time of grain axis of maximum moment of inertia around the radiation direction, $\tau_{k}$, if grains thermally rotate, i.e., $\omega \sim \omega_{T}$ (see Table \ref{tab:ZC}). However, $\tau_{k}$ becomes longer than $\tau_{B}$ as the grain rotates suprathermally under the action of strong RATs. Thereby, the Larmor precession around the magnetic field is still faster than the precession around the radiation direction, and the alignment axis is directed along the magnetic field. Superparamagnetic grains should always be aligned with respect to the magnetic fields.

\subsubsection{Circular Polarization of Zodiacal light}

Below we calculate the CP degree arising from the scattering of sunlight by aligned Zodiacal dust. Each line of sight is characterized by an elongation angle $\epsilon$. The total intensity of scattered light is obtained by integrating Equation (\ref{eq:dI}) along the line of sight and over the entire grain size distribution:
\bea
I(\epsilon)=\int \int_{a_{\min}}^{a_{\max}}\Delta I(\Rv,\rv) da dr,\label{eq:Isca}
\ena
where we adopt $a_{\min}=0.1\mum$ and $a_{\max}=10\mum$.

Similarly, the intensity of circularly polarized light is obtained by integrating Equation (\ref{eq:dV}) over the line of sight and the range of grains which are aligned:
\bea
V(\epsilon)=\int \int_{a_{\ali}}^{a_{\max}}\Delta V(\Rv,\rv) da dr.\label{eq:Vpol}
\ena

The CP degree then becomes $q(\epsilon)=V/I$. The geometry of calculations is shown in Figure \ref{fig:sketch}. The integrals above can be numerically solved. The vector term in Equation (\ref{eq:dV}) can be rewritten as
\bea
[\hat{\Rv}\times \hat{\rv}].{\bf h}=\sin\psi\sin\theta_{B},\\
\hat{\Rv}.{\bf h}=\cos\theta_{B}\cos\alpha,
\ena
where $\theta_{B}$ is the angle between $\Bv$ and the ecliptic plane, $\psi$ is the scattering angle between $\Rv$ and $\rv$, and $\alpha$ with $\tan\alpha=R/R_{0}$ is the angle between the projection of $\Bv$ onto the ecliptic plane and the $\Rv$ direction (see Figure \ref{fig:sketch}).

\begin{table}
\centering
\caption{Model parameters and characteristic timescales for Zodiacal dust}\label{tab:ZC}
\begin{tabular}{l l l} \hline\hline\\
Parameters & Notations & Values\cr
\hline\\
Gas density & $n_{\H}$& $ 5\left(\frac{R}{\AU}\right)^{\alpha}\cm^{-3}$\cr
Radiation density & $u_{\rad}$ &$2\times 10^{-4}\erg\cm^{-3}$\cr
Gas temperature & $T_{\gas}$& $10^{4}\K$\cr
Gas velocity at 1 $\AU$ & $v$ & $3-400\km \s^{-1}$\cr
Magnetic field 1 $\AU$ & $B$ & $10-50 \mu$G\cr
Gas damping & $\tau_{\gas}$ &$4.4\times10^{4}a_{-5}\yr$\cr
IR damping & $\tau_{\rm IR}$&$0.2a_{-5}^{2}\yr$\cr
Davis-Greenstein alignment & $\tau_{\DG} $ &$3.2\times 10^{3}a_{-5}^{2}\yr$\cr
{Larmor precession}    & $\tau_{B}$ & $8\times 10^{-2}a_{-5}^{2}\left(50\mu {\rm G}/B\right)\yr$\cr
{Precession around radiation} & $\tau_{k}$ & $ 10^{-3}a_{-5}^{1/2}\left(\omega/10^{2}\omega_{T}\right)\yr$\cr
{Mechanical alignment} & $\tau_{\rm Gold}$ & $4a_{-5}\yr$\cr
\cr
\hline\\
\end{tabular}
\end{table}

\begin{figure}
\includegraphics[width=0.45\textwidth]{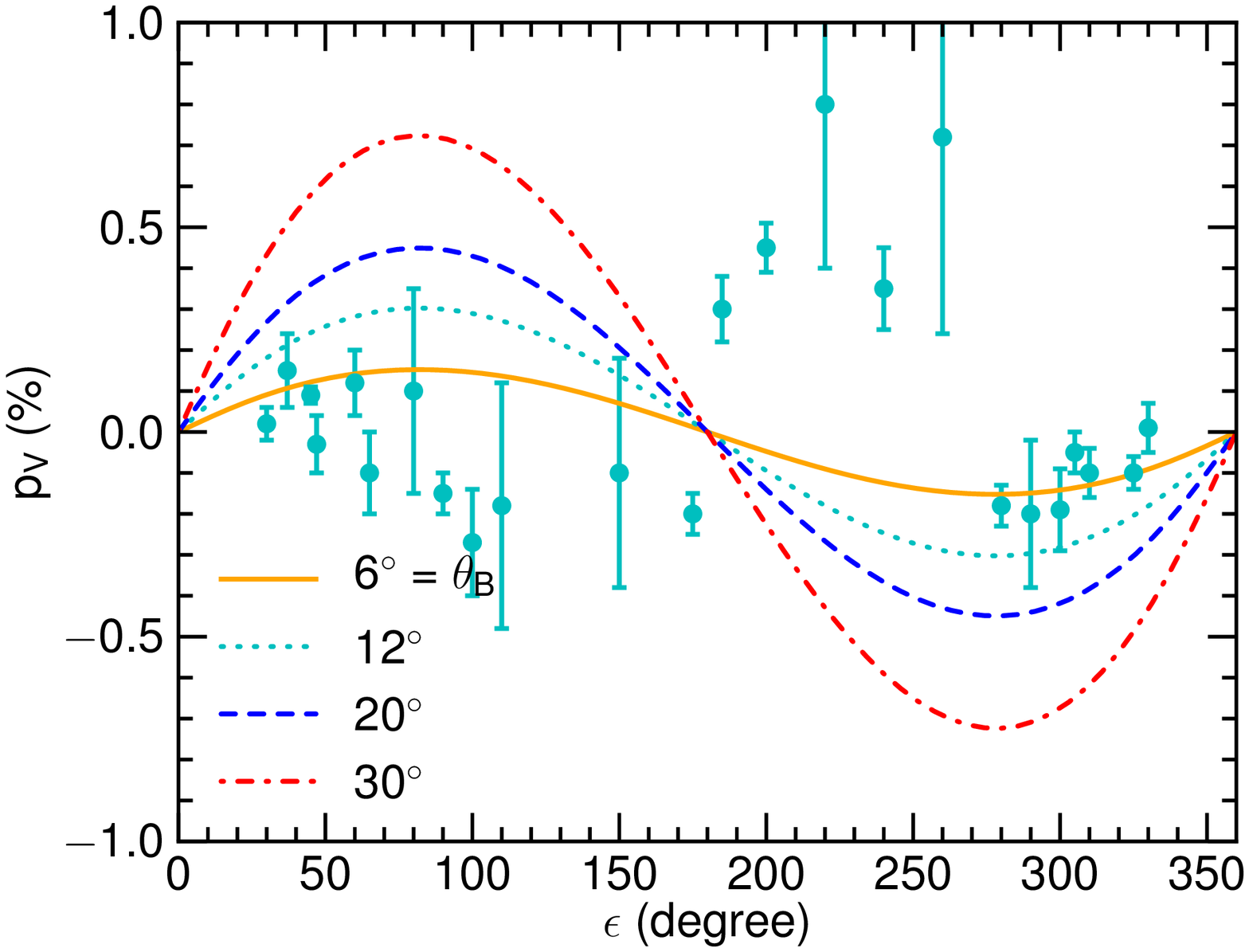}
\includegraphics[width=0.45\textwidth]{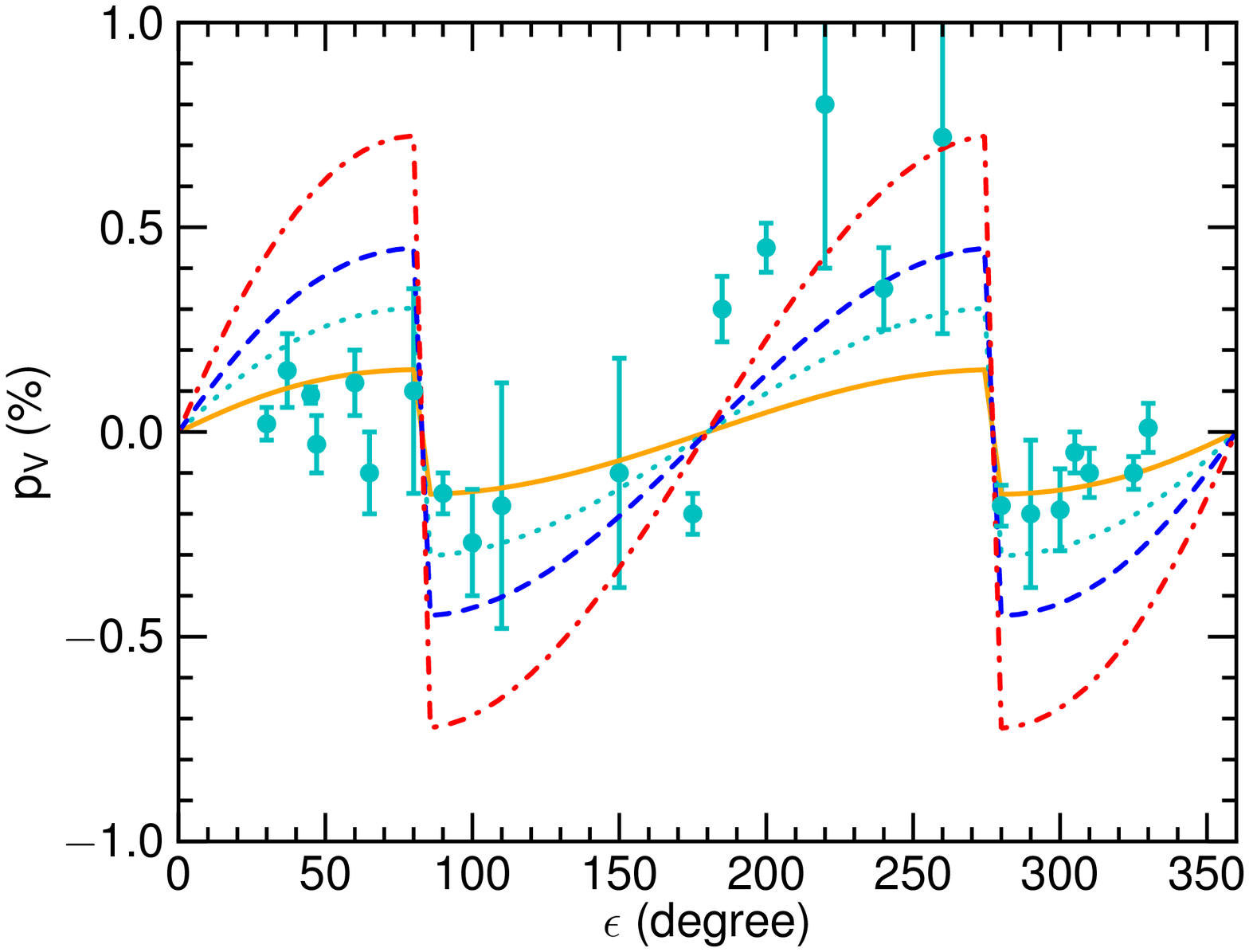}
\caption{CP at the optical wavelength $q_{V}$ as a function of the elongation angle $\epsilon$ for the different angles $\theta_{B}$ between the magnetic field and the ecliptic plane. Here $\epsilon=90^{\circ}$ indicates the line of sight perpendicular to the Sun-Earth direction. In the upper panel, the magnetic field is assumed to be spiral with inclusion of a vertical component parallel to the north ecliptic pole. In the lower panel, the vertical component of the magnetic field is reversed for $\epsilon<90^{\circ}$ and $\epsilon>270^{\circ}$. The observational data obtained from Wolstencroft \& Kemp (1972) are shown in filled circles. Strong polarization observed for $\epsilon=250-280^{\circ}$ may reveal a higher inclination of the magnetic field from the ecliptic plane or local enhancement of grain alignment.}
\label{fig:pV_Zod}
\end{figure}

The upper panel in Figure \ref{fig:pV_Zod} shows the degree of CP, $q_{V}$, as a function of the elongation angle $\epsilon$ for the different angles of $\Bv$ with the ecliptic plane, $\theta_{B}$. One can see that the predicted $q_{V}$ is comparable to the observational data. In addition, $q_{V}$ increases with the increasing $\theta_{B}$, i.e., $q_{V}$ increases with the increasing vertical magnetic field. But, the variation of $q_{V}$ with $\epsilon$ does not agree well with the data. 

To obtain a better agreement with the observational data, we change the direction of the vertical magnetic field for $\epsilon=90^{\circ}-270^{\circ}$. Our obtained results are shown in the lower panel. One can see that this case gives rise to a better agreement with the observational data.

\subsection{Cometary comae}\label{sec:comet}
\subsubsection{Observational studies for polarization from cometary dust}

Numerous observations (see \citealt{2007Icar..186..317R} and references therein) have showed that scattered light from cometary comae exhibits both linear polarization and CP. Multiple scattering by irregular dust grains is frequently appealed to explain the linear polarization from binary systems and circumstellar disks in which the polarization direction is perpendicular to the scattering plane (i.e., normal polarization; see \citealt{Bastien:1988te}). In particular, some polarimetric observations reveal an anomalous feature, namely, polarization vectors lying in the scattering plane, which is unexpected from the multiple scattering (\citealt{1971A&A....14...90C}). Moreover, the magnitude of such an anomalous polarization is comparable to the normal polarization. Scattering by aligned dust grains was suggested as a cause of the anomalous polarization by \cite{Dolginov:1976p2480}, and the existence of aligned grains in coma was indeed identified through starlight polarization during occultation (\citealt{1994Icar..108...81R}). However, underlying mechanism for the alignment of cometary dust and quantitative modeling of polarization are not yet available. 

Recent observations of CP throughout the coma of Comet C/1999 S4 were reported in \cite{2007Icar..186..317R}. Their data show that the CP exhibits some systematic variation along the cuts through the coma and nucleus and is equal to zero at the nucleus. In this section, we intend to model the grain alignment in a typical coma and predict the CP of sunlight scattered by aligned grains.

\subsubsection{Model Setup}
The cometary coma is assumed to be spherical in which gas and dust are being produced continuously from an icy nucleus (consisting of mostly frozen H$_{2}$0, NH3, CH4, etc.) due to solar radiation. Since the nucleus presumes to be heated symmetrically by sunlight because its rotation period ($\sim 10^{4}\s$) is much shorter than the orbital period ($\sim 10^{6}\s$), gas and dust are expanding symmetrically in the radial direction.\footnote{A realistic geometry of the inner coma should be a fan-like due to the effect of radiative heating is stronger in the sunward direction. Of course, in the outer region of the coma where the radiative pressure becomes dominant, the asymmetry is elongated in the tailward direction.} 

Let $Q_{\gas}$ be the rate of mass production by the cometary nucleus and $v_{\gas}$ be the expansion velocity of gas. The gas mass density at distance $r$ from the nucleus can be given by (\citealt{Haser:1957uea}):
\bea
\rho_{\gas}=\frac{Q_{\gas}}{4\pi v_{\gas} 
r^{2}}\exp\left(-\frac{r}{L_{g}}\right),\label{eq:ngas_coma}
\ena
where $dM=\rho_{\gas}4\pi r^{2}v_{\gas}dt=Q_{\gas}dt$ is the mass produced during the time interval $dt$, $L_{g}$ is the ionization length scale, which is between $1-2\times 10^{6}\km$ (see \citealt{1991JGR....96.7731L}). Above, the exponential term describes the decay of gas, and the subdominant effect of solar radiative pressure on the expanding gas is disregarded. For a coma with the radius $r\ll L_{g}$, we can ignore the exponential term in Equation (\ref{eq:ngas_coma}). Physical parameters for a coma are listed in Table \ref{tab:comet}. The gas production rate $Q_{\gas}=9\times 10^{4}\g\s^{-1}$ is assumed (\citealt{2007Icar..186..317R}).

\begin{table}
\centering
\caption{Model parameters for a cometary coma}\label{tab:comet}
\begin{tabular}{l l}\hline\hline\\
Parameters & Values \cr
\hline\\
{Radius of nucleus}& {$1\km$}\cr
{Star temperature}& {$T_{\eff}=5800\K$}\cr
{Star luminosity}& {$L=L_{\odot}$}\cr
{Star radius}& {$R=R_{\odot}$}\cr
{Gas density}& {$n=n_{0}\left(R_{n}/r\right)^{2}$} $^{\rm a}$\cr
{Gas temperature}& {$T_{\gas}=300\K$}\cr
{Expansion velocity}& {$v_{\gas}=1 \km\s^{-1}$}\cr
\cr
\hline\\
\multicolumn{2}{l}{$^{\rm a}$ Here $n_{0}=4\times 10^{12} \cm^{-3}$ at $R_{n}=1$ km}
\end{tabular}
\end{table}

Since the mean free path of gas molecules in the inner coma ($n\sim 10^{10}-10^{12}\cm^{-3}$) is much smaller than grains size (high density), dust near the nucleus is accelerated by molecular drag force and dragged away by the gas flow. Dust is decoupled from the gas at some distance from the nucleus and achieves terminal velocity at a distance of tens the nucleus radius, which is between $20$ and $100\km$ (see e.g., \citealt{1968ApJ...154..327F}). 

Grains are electrically charged as a result of photoemission by solar radiation and collisions with electrons and ions. For the inner coma where the collisional charging dominates, grains are expected to be negatively charged with their potential from $U=0$ to $-1$V. For the outer regions where the photoemission dominates, grains are positively charged with the potential from $U=1$ to $10$V (\citealt{1983A&A...121...10W}).
  
\subsubsection{Grain alignment in cometary comae}\label{sec55}
The problem of grain alignment in cometary comae appears to be a puzzle because of their extreme conditions. Unlike ZC, in which grains can be aligned by RATs in the direction of the IMF, in comae, the alignment with magnetic fields is apparently impossible because solar winds that carry magnetic fields cannot penetrate into the dense region of coma with radius less than a few thousand kilometers from the comet nucleus.

The grain alignment cannot occur with respect to the solar radiation direction either because this alignment direction lies in the scattering plane, which produces zero CP (see Equation (\ref{eq:dV})).

Dust grains in the coma may be aligned by the gaseous outflow via the Gold mechanism (\citealt{Gold:1952p5848}; \citealt{1994MNRAS.268..713L}). Assuming the gaseous outflow velocity $v_{\rm flow}=10^{5}\cm\s^{-1}$, one can estimate the alignment timescale by the Gold mechanism using Equation (\ref{eq:tauGold}):
\bea
\tau_{\rm Gold}=4.7\times 10^{6}a_{-5}\left(\frac{n_{\H}}{10^{10}\cm^{-3}}\right)^{-1}
\left(\frac{v_{\rm flow}}{10^{5}\cm\s^{-1}}\right)^{-3} \s,~~~
\ena
and the rotational damping time is equal to
\bea
\tau_{\drag}=10^{4}a_{-5}\left(\frac{n_{\H}}{10^{10}\cm^{-3}}\right)^{-1}\left(\frac{300\K}{T_{\gas}}\right)\s,
\ena
where $F_{\rm IR}\ll 1$ is disregarded. Since $\tau_{\rm Gold}>>\tau_{\drag}$, the Gold alignment is inefficient for aligning cometary grains.

In Equation (\ref{eq:tauE}) we show that the grain dipole can rapidly precess around electric fields. It is noted that the static electric field can exist for grains at rest in plasma. For instance, the electric field is produced by charged particles arising from the photoionization of gaseous atoms/molecules by solar UV radiation.\footnote{The ionization degree in the coma has been measured for the Comet Halley by \cite{1991JGR....96.7731L}.} Since the expansion direction is radial, the electric field is likely directed along the radial direction.

\begin{figure}
\includegraphics[width=0.45\textwidth]{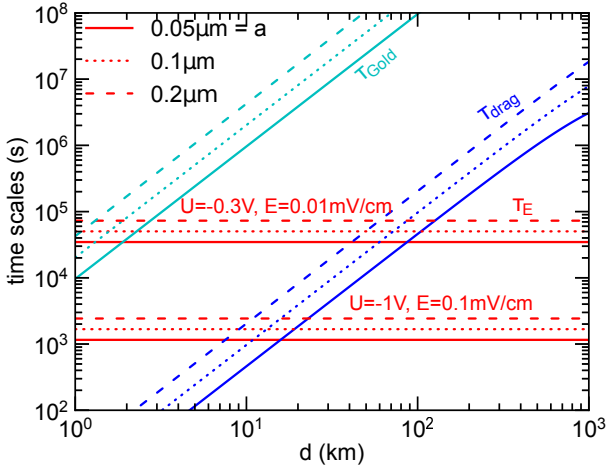}
\caption{Characteristic timescales, $\tau_{E}, \tau_{\drag}$, and $\tau_{\rm Gold}$ as a function of the distance $d$ from the nucleus for different grain sizes. Two realizations of $U$ and $E$ are considered. For $d>50\km$, the electric field is the axis of grain alignment because $\tau_{\drag}>\tau_{E}$. Grain center of mass is assumed to be at rest.}
\label{fig:tauE-coma}
\end{figure}

Figure \ref{fig:tauE-coma} shows the precession time of the grain dipole around the electric field, $\tau_{E}$, compared to $\tau_{\drag}$ and $\tau_{\rm Gold}$. We consider two cases: $U=-0.3$ V, $E=10^{-5}$V/cm and $U=-1$V, $E=10^{-4}$V/cm. It is noted that the measurements by Voyager 1 and 2 for the Halley comet indicate that the electric field $E\sim 10^{-4}V/cm$ is expected (\citealt{1991JGR....96.7731L}). For the first case, the figure shows that $\tau_{\drag}\ge \tau_{E}$ for $d<50\km$, thus, the electric field does not play the role of axis of alignment. Beyond $d=50\km$, $\tau_{E}<\tau_{\drag}$, and the electric field becomes the axis of grain alignment. The distance at which $\tau_{E}\sim \tau_{\gas}$ is lower for the second case.

To study the alignment of grains by RATs in coma, we assume that the coma is a sphere of radius $R$, with the gas density decreasing with the distance as given by Equation (\ref{eq:ngas_coma}). In general, the coma radius $R$ varies with the distance from the comet to the Sun due to the variation of solar heating. For the coma with $R=10^{4}$km, the column density corresponds to $N_{\gas}\approx 10^{18}\cm^{-2}$ using the function $n(r)$. The optical depth can be estimated as $\tau=N_{\gas}\sigma_{\ext}\approx 10^{18}\times 10^{-21}=10^{-2}\ll 1$. Thus, the radiation is uniform throughout the optically thin coma. Taking the radiation intensity from the Sun and the rotational damping due to gas and IR emission, we can calculate the critical size of aligned grains $a_{\ali}$ (see \S \ref{sec:mod}).

\begin{figure}
\includegraphics[width=0.45\textwidth]{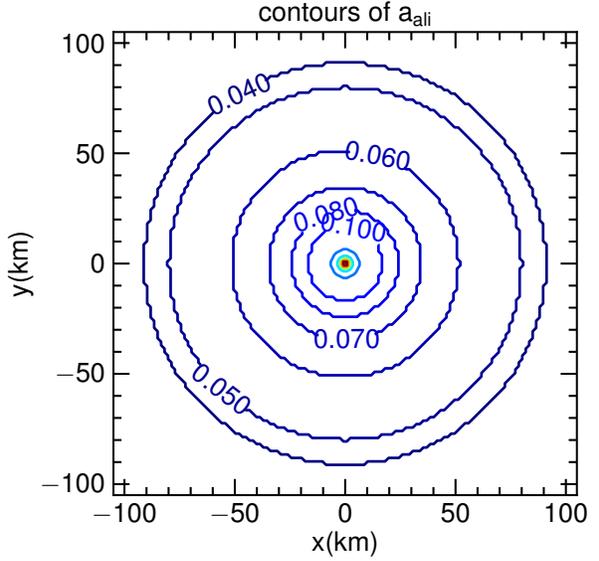}
\caption{Contours of the critical size of aligned grains $a_{\ali}$ by RATs for the inner region of a cometary coma at a distance $R_{\rm helio}=1\AU$ from the Sun. The contours are nearly concentric because the coma is optically thin with the radial dependence of the gas density. Grain center of mass is assumed to be at rest.}
\label{fig:aali-Comet}
\end{figure}

Figure \ref{fig:aali-Comet} shows the contours of $a_{\ali}$ obtained from our calculations, assuming that grains are at rest relative to the nucleus. It can be seen that the contours of $a_{\ali}$ are nearly concentric, indicating that the radiation energy density is uniform in the coma. This feature arises from the fact that the coma is optically thin due to its small size, i.e., $\tau\ll 1$, such that the radiation energy density $u_{\rad}$ is uniform throughout the coma. As a result, $a_{\ali}$ is governed by the rotational damping, which depends on the gas density. Within a radius $d <20$ km of the comet nucleus, the RAT alignment is still efficient, but only for larger grains ($a>0.1\mum$). In the outer layers of radius $d >20$ km, smaller grains can efficiently be aligned by RATs. Noting that for the inner region $d<50\km$, the axis of alignment is not directed along the electric field because $\tau_{\drag}<\tau_{E}$ (see Figure \ref{fig:tauE-coma}).

Although grains within the radius $d<100\km$ can be aligned by RATs (see Figure \ref{fig:aali-Comet}), it should be kept in mind that the alignment time must be shorter than the lifetime of cometary dust to produce the observed polarization. For an escaping velocity $v_{d}\sim 0.1\km \s^{-1}$ (\citealt{1992Icar...97...85W}), the grain will leave the coma after $t_{out}=R/v_{d}\sim 10^{3}\left(R/10^{2}\km\right)\s$. Therefore, the precession time $\tau_{E}$ is longer than the dust lifetime for $R<100\km$, i.e., grains are randomly oriented in this region. Grains seem to be aligned with the electric field in the region $R>100\km$.

\subsubsection{Circular Polarization from Cometary Comae}
The position of a comet is determined by its distance to the Sun, so-called heliodistance $R_{\rm helio}$, and its distance to earth, so-called geodistance $R_{\rm geo}$. The phase angle of the comet is equal to the angle between $R_{\rm helio}$ and $R_{\rm geo}$. Consider an volume element $d\Gamma$ of dust in the coma, which is characterized by the radius vector $\rv_{c}$ from the nucleus. Since the radius of the cometary coma is much smaller than $R_{\rm helio}$ and $R_{\rm geo}$, in Equation (\ref{eq:dI}) and (\ref{eq:dV}) we can replace $R \approx R_{\rm helio}$ and $r\approx R_{\rm geo}$.  

The Stokes V parameter can be rewritten as
\bea
\delta V\propto \left([\Rv\times \rv].\hv\right)\left(\Rv.\hv\right).
\ena

We can represent $\Rv$ and $\rv$ as
\bea
\Rv=R_{\rm helio}+\rv_{c},
\rv	= R_{\rm geo}+\rv_{c}.
\ena

The electric field vector, also the axis of alignment, is assumed to be along the radial direction, i.e. $\hv\equiv \rhat_{c}$. From the above equations we obtain
\bea
[\Rv\times \rv].\hv=\left[\Rv_{\rm helio}\times \Rv_{\rm geo}\right].\rhat_{c}=\sin\alpha\cos\theta,
\ena
and
\bea
\Rv.\hv&=&(\Rv_{\rm helio}+\rv_{c}).\rhat_{c},\nonumber\\
&\approx& \Rv_{\rm helio}.\rhat_{c}=\sin\theta\sin(\psi+\phi_{\odot}).
\ena

Thus,
\bea
\delta V\propto \left(\sin\alpha \cos\theta\right)\left[\sin\theta\sin(\psi+\phi_{\odot})\right],
\ena
where $\theta, \psi$ are the polar angle and azimuthal angle of $\rv_{c}$ in the $\xhat\yhat\zhat$ coordinate system, $\alpha$ is the phase angle between $\Rv_{\rm helio}$ and $\Rv_{\rm geo}$, and $\phi_{\odot}$ is the angle between the Comet-Sun direction and Comet-earth direction (i.e., $\xhat$ axis). 

We calculate $\delta I$ and $\delta V$ for the solar radiation scattered towards the observer by a volume element of comet dust and then integrate along the line of sight ($\xhat$ direction) to obtain $I(y,z)$ and $V(y,z)$ in the sky plane $\yhat\zhat$. Hence, the CP is given by $q(y,z)=V/I$.

\begin{figure}
\includegraphics[width=0.45\textwidth]{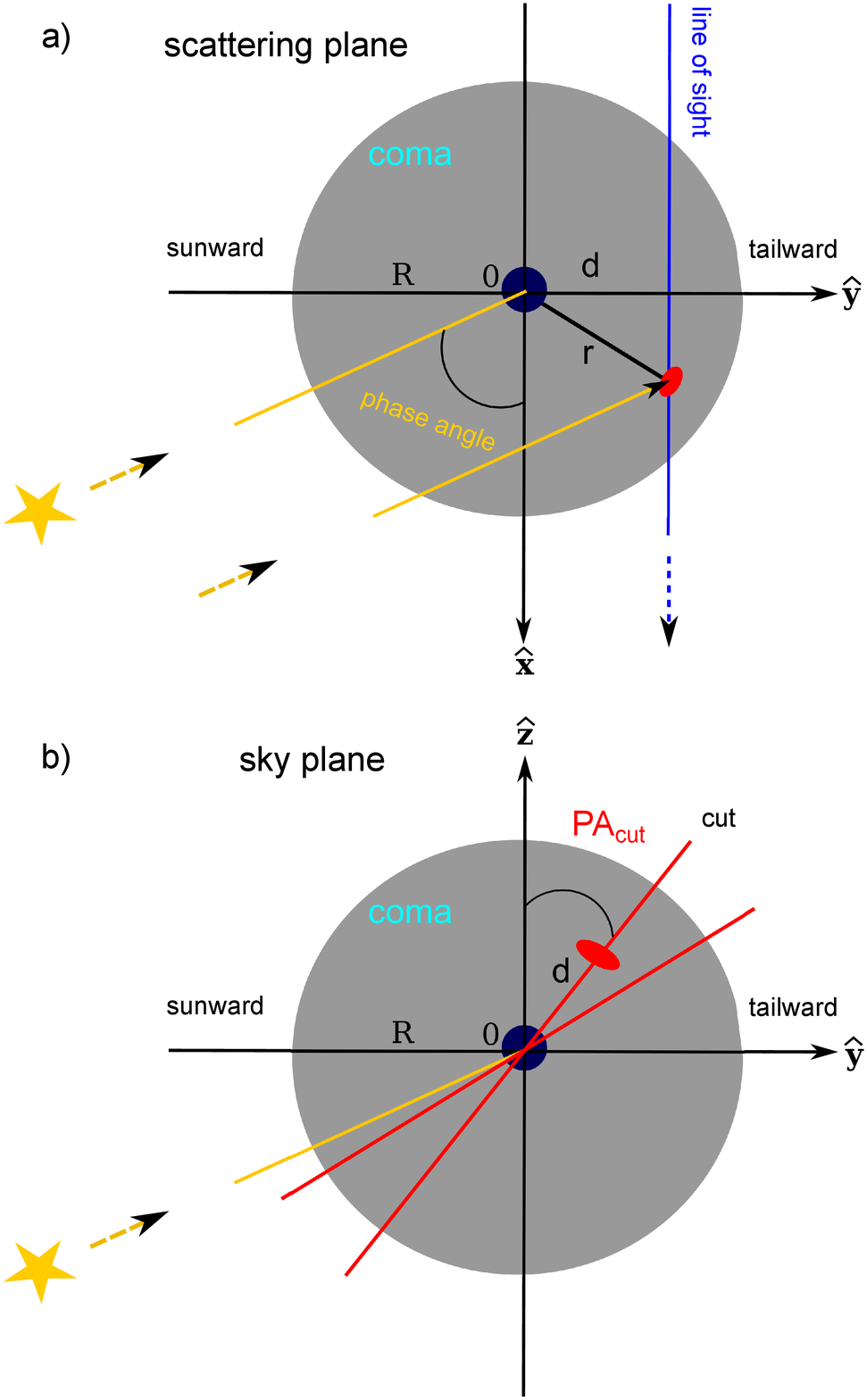}
\caption{(a) Schematic of a cometary coma used for calculations of CP. Dust grains are aligned with respect to the electric field in the radial direction. The line of sight is chosen along the $\xhat$ direction, and $d$ is the projected distance of dust grains onto the sky plane $\yhat\zhat$ ($\zhat$ axis perpendicular to the scattering $\xhat\yhat$ plane) to the comet nucleus. (b) Cometary coma in the sky plane. Scattered light is observed along the cuts that go through the coma and nucleus. Each cut is determined by the position angle, PA$_{\rm cut}$.}
\label{fig:comets}
\end{figure}

Grains in coma may have irregular shape or aggregate of fluffy grains, and exist in a wide range from submicron to centimeter size (see \citealt{2004come.book..577K} and references therein). For the optical wavelength under interest, the scattering intensity $I$ is dominated by the $a\sim \lambda$ grains and the contribution by large grains with $2\pi a \gg \lambda$ may be negligible. In addition, the intensity $V$ of circularly polarized scattered light is determined by the absorption through the imaginary part of the dielectric function, while the absorption is minor for $\lambda \le a$. \cite{2000Icar..143..338J} and \cite{2001ApJ...549..635M} suggested that very small grains can be present in comae. For our calculations, grains are assumed to have oblate spheroidal shape and their size follows a power law distribution, i.e., $n(a)\propto a^{-\alpha}$ with $\alpha=3.5$ with the cutoffs $a_{\min}=0.1\mum$ and $a_{\max}= 10\mum$, which is consistent with observations for the Comet Halley (see \citealt{1986Natur.321..338M}; \citealt{1987A&A...187..719M}; \citealt{2004come.book..265H}).\footnote{Grain size distribution can be probed by modeling thermal emission from the coma with the use of observations. Studies by \cite{1998A&A...338..364L} show that a steeper power law with $\alpha=1.5-2.5$ is expected.} Figure \ref{fig:comets} shows the schematic illustration of our calculations consisting of the scattering plane (upper) and the sky plane (lower). To compare with observations, the cuts through the comet nucleus and coma (red lines) are indicated. Along each cut, we calculate the polarization at difference distances $d$ from the nucleus. 

\begin{figure}
\includegraphics[width=0.45\textwidth]{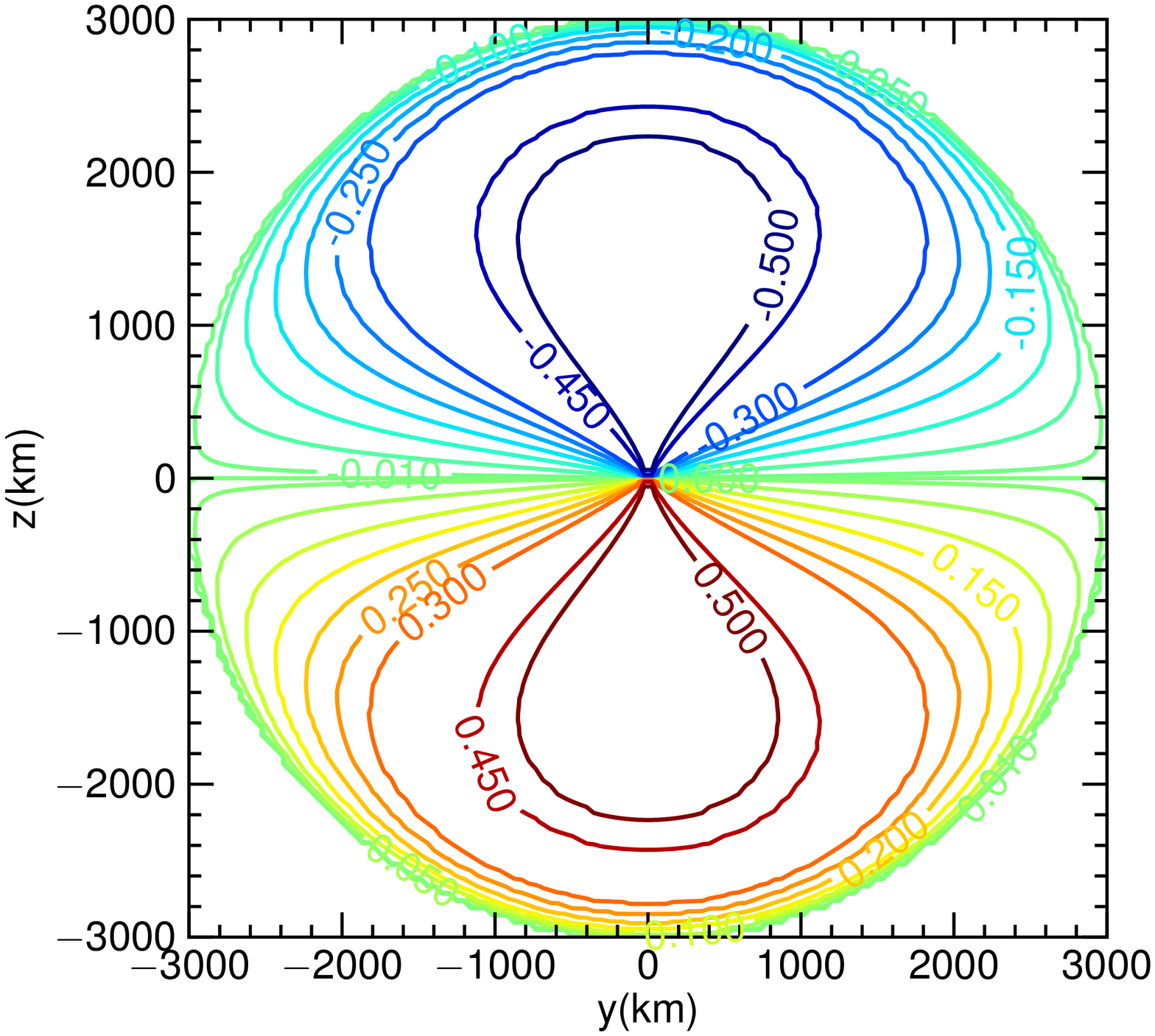}
\includegraphics[width=0.45\textwidth]{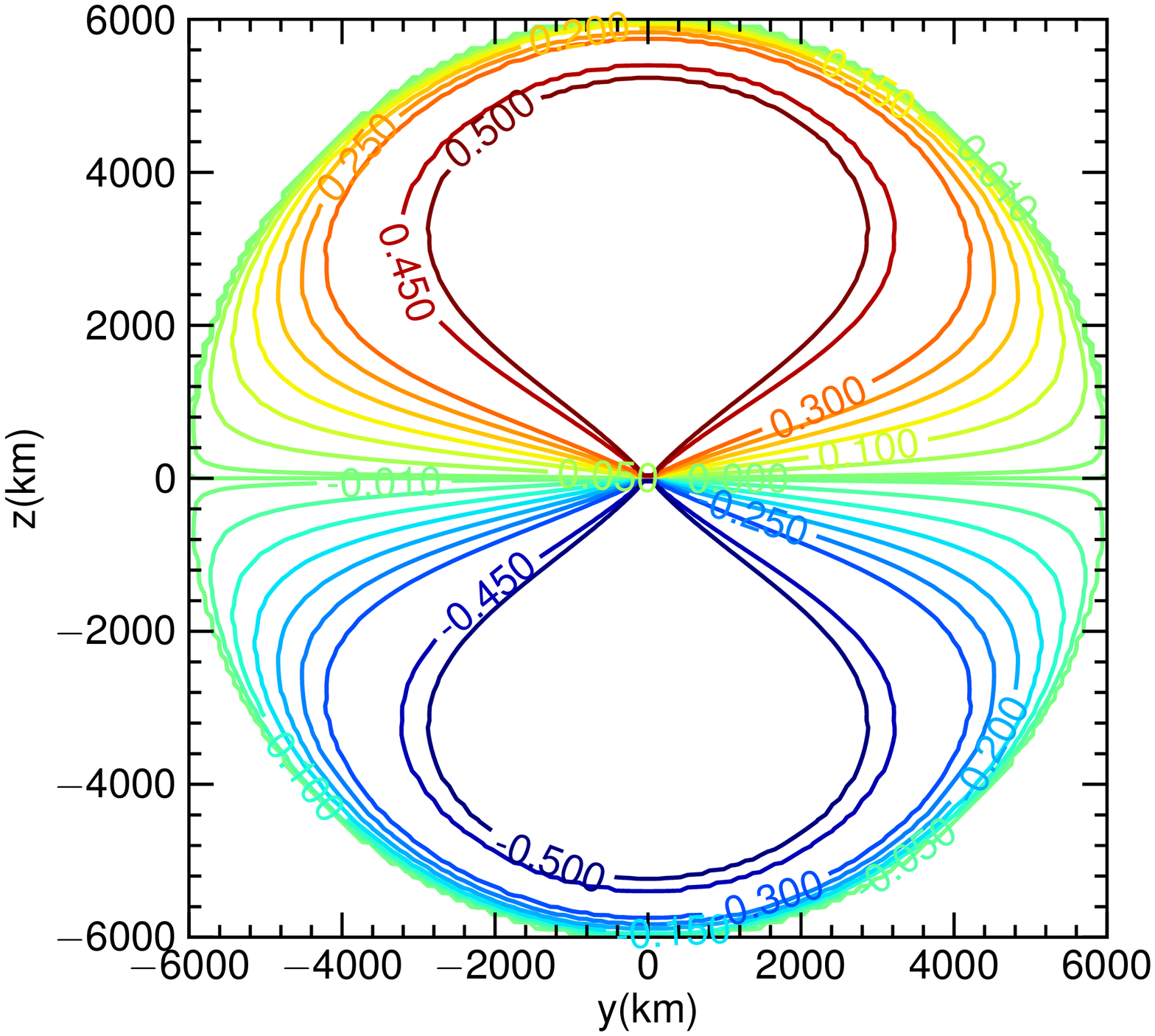}
\caption{Contours of CP at the optical wavelength, $q_{V}(y,z)$, in the sky plane for two cases of phase angle $\alpha=68.4^{\circ}$ (upper) and $121.3^{\circ}$ (lower). In the second case the comet is closer to the Sun and its radius becomes larger. }
\label{fig:pVcomet}
\end{figure}

Figure \ref{fig:pVcomet} shows the contours of CP, $q_{V}$, in the sky plane predicted for two positions of the comet determined by the phase angles $\alpha=68.4^{\circ}$ (upper) and $121.3^{\circ}$ (lower). The heliodistance $R_{\rm helio}=0.89\AU$ and $0.77\AU$, and the geodistance $R_{\rm geo}=0.92\AU$ and $0.38\AU$, respectively.
One can see that the contours of $q_{V}$ are anti-symmetric with respect to the $\yhat$ axis, which is a direct result of the antisymmetry of the electric field directed in the radial direction.

\begin{figure}
\includegraphics[width=0.45\textwidth]{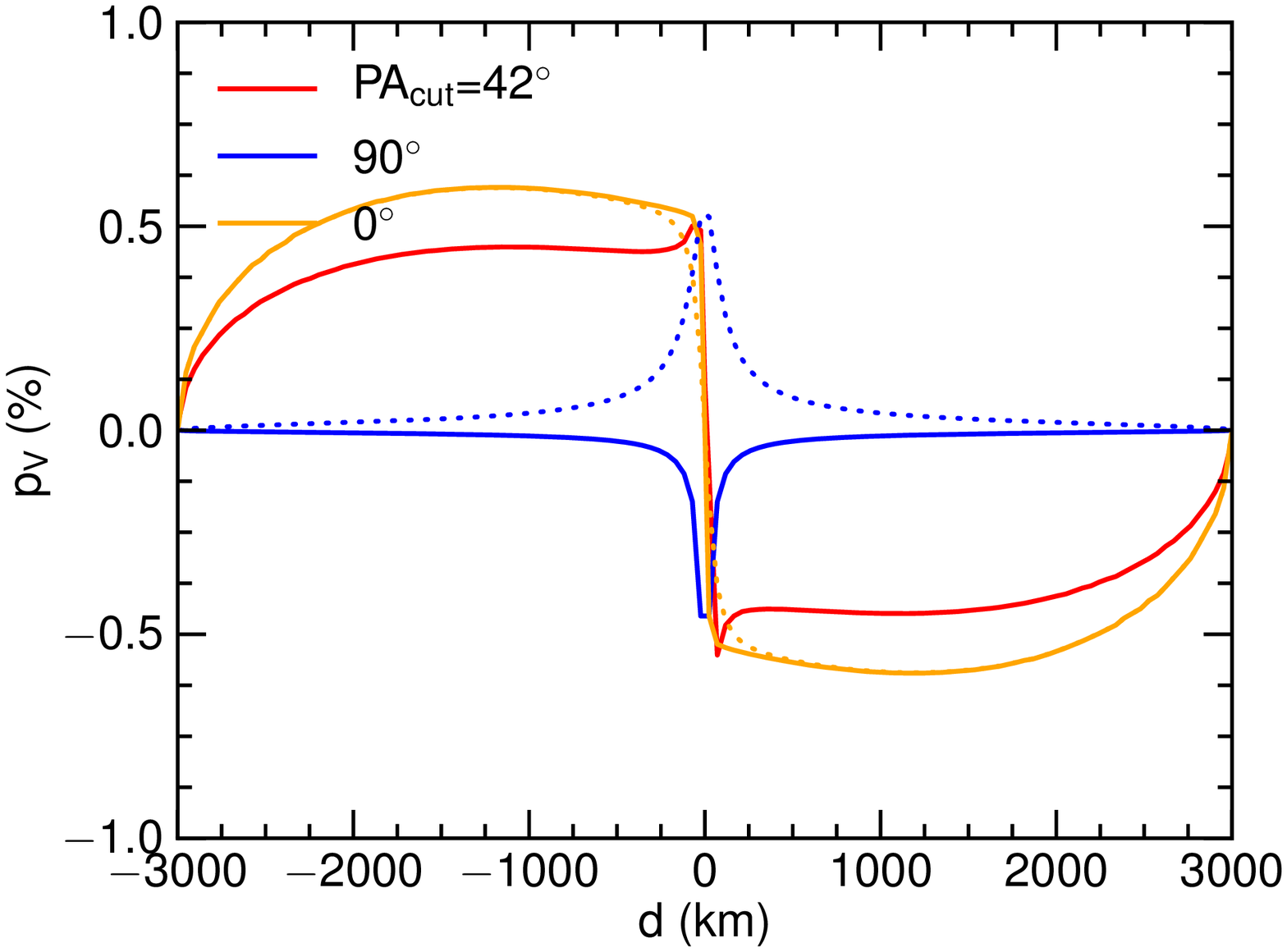}
\includegraphics[width=0.45\textwidth]{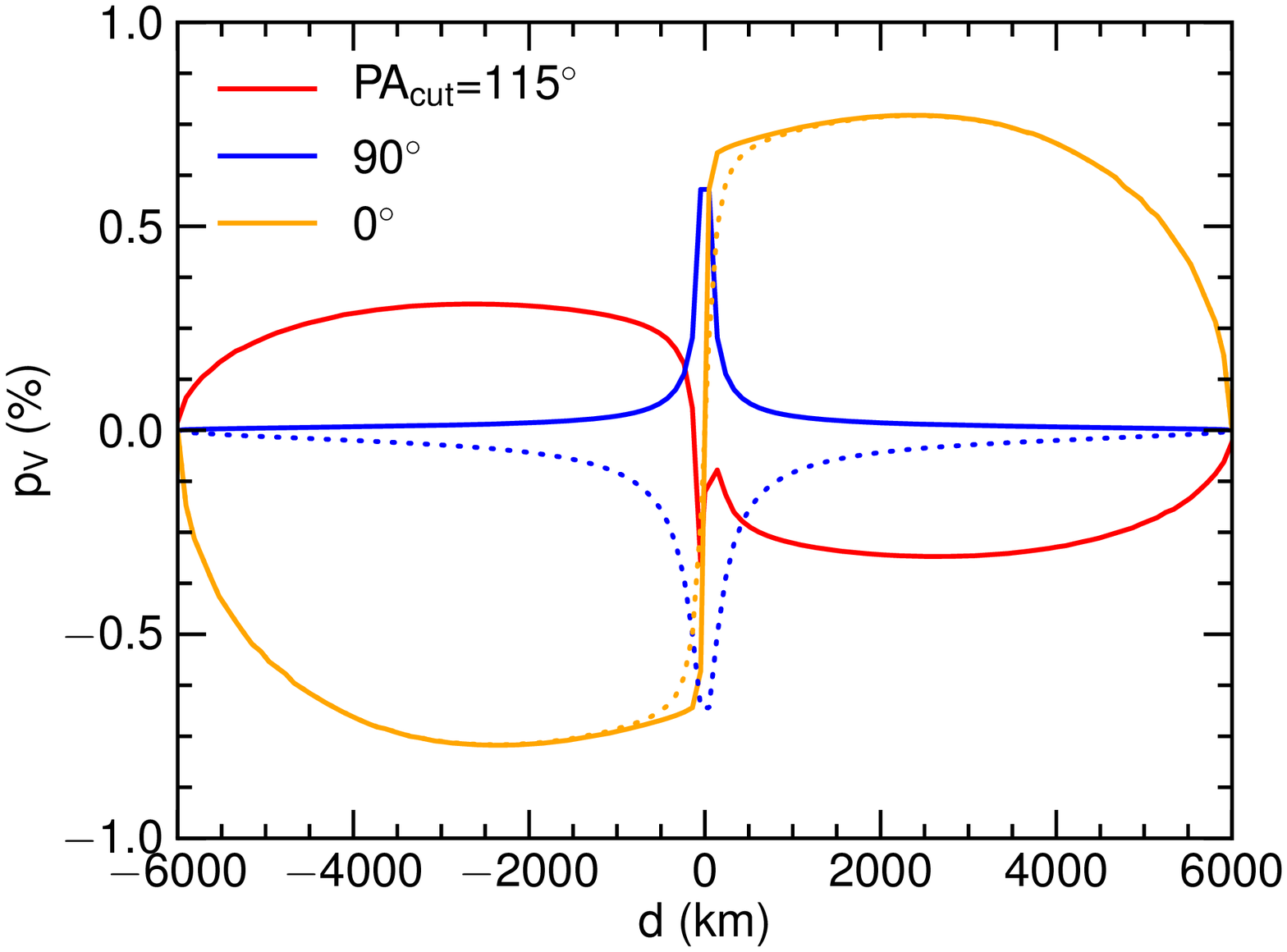}
\caption{Circular polarization $q_{V}$ as a function of the distance from the comet nuclei along the different cuts with position angle PA$_{\rm cut}=0^{\circ}$ (cut along $\zhat$ axis), $30^{\circ}$ and $90^{\circ}$ (cut perpendicular to $\zhat$ axis). Upper and low panels show the results for two phase angles $\alpha=68.4^{\circ}$ and $121.3^{\circ}$, respectively. Solid lines show the cut through both the coma and nucleus, whereas dotted lines show the results for cuts through the coma and off-center by $20$ km.}
\label{fig:pVcuts}
\end{figure}

Figure \ref{fig:pVcuts} shows the variation of $q_{V}$ along a cut in the sunward ($d<0$) and tailward ($d>0$) directions for the different cuts determined by the position angle PA$_{\rm cut}$. Upper and lower panels show the results for the same phase angles as those in Figure \ref{fig:pVcomet}. For the cuts with $PA_{\rm cut}\ne 90^{\circ}$ and through both coma and nucleus (solid line), $q_{V}=0$ at $d=0$. The absolute value increases with distance first an decreases to zero at the coma boundary. The sign of $q_{V}$ is opposite for the sunward and tailward directions and changes it sign at $d=0$. Moreover, the variation of $q_{V}$ is very sharp for the cut through the nucleus (see solid line), whereas the variation is smoother for the cut at some distance from the nucleus (see dotted line). For the cuts of $PA_{\rm cut}=90^{\circ}$, $q_{V}$ is nonzero only for the cuts not through the nucleus. 

Interestingly, these systematic variations are consistent with observations by \cite{2007Icar..186..317R}. For example, they found that $q_{V}$ is positive for June 28, 2000 when the cuts do not go through the center. For other dates with the cuts through the nucleus, $q_{V}$ is very small at $d=0$, which is consistent with our predictions.

\begin{figure}
\includegraphics[width=0.45\textwidth]{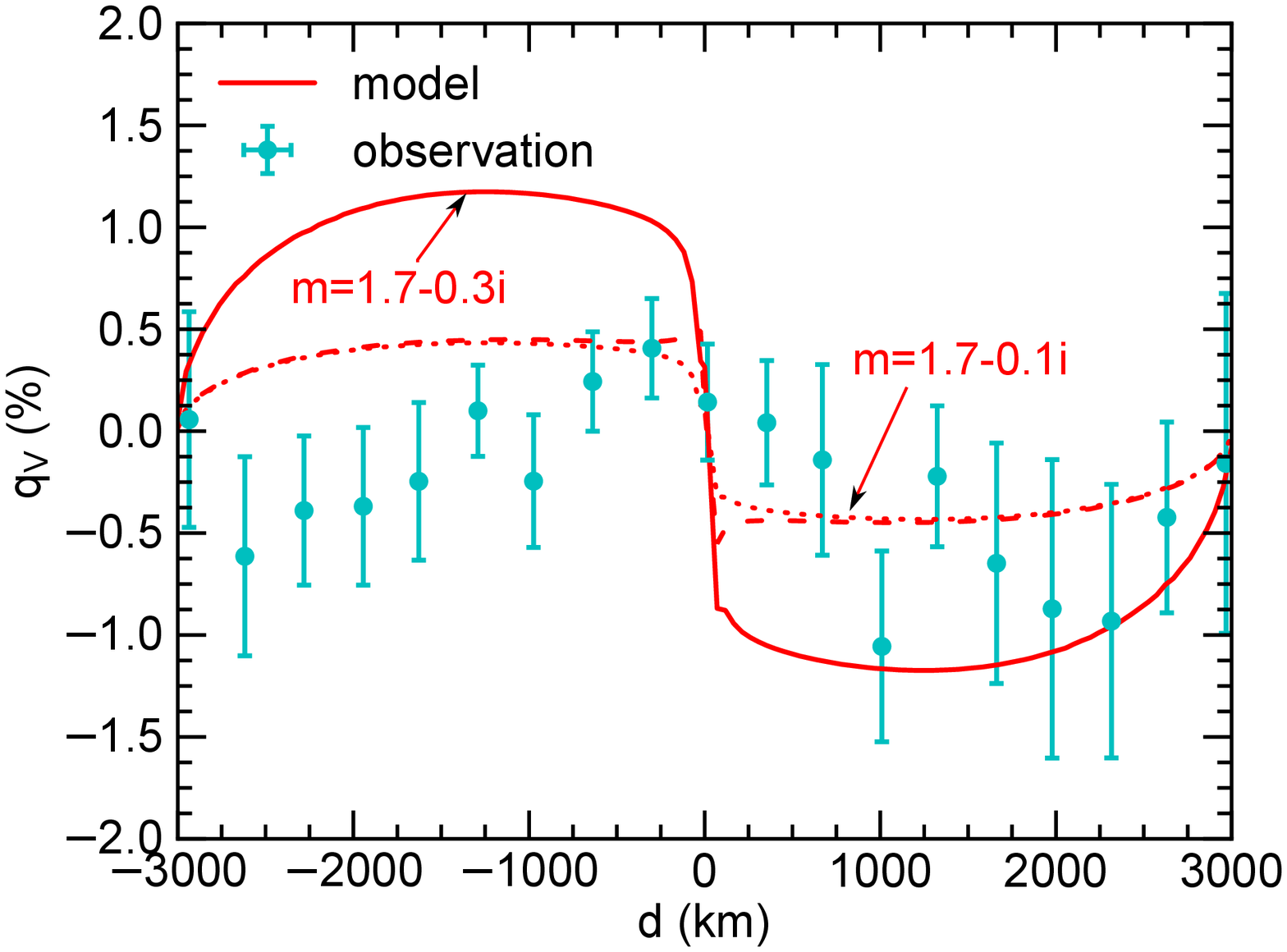}
\caption{Comparison of our model with the observed CP of Comet C/1999 S4 on July 2, 2000 by Rosenbush et al. (2007). Dotted (dashed) and solid lines show the CP assuming the complex refractive index $m=1.7-0.1i$ and $m=1.7-0.3i$ for silicate grains. Two cuts, one through the nucleus (dashed line) and one off-center by 20 km (dotted line), are considered. Model is in good agreement with the observational data for $d<1000\km$.}
\label{fig:pVcomet_obs}
\end{figure}

In Figure \ref{fig:pVcomet_obs} we compare our predictions with the observational data from \cite{2007Icar..186..317R} for a particular comet position $\alpha=68.4^{\circ}$. There, we show the predictions for a cut with $PA_{\rm cut}=42.2^{\circ}$ and for the imaginary part of the refractive index Im(m)=-0.1 (dashed-dot line) and Im(m)=-0.3 (solid line). The latter increases $q_{V}$ by a factor of $2.7$ due to the increase of the polarizability. 

From Figure \ref{fig:pVcomet_obs}, it can be seen that, within the region $|d|<1000\km$, the model is in good agreement with the observational data. Indeed, both the model and observational data show the anti-symmetry of $q_{V}$ through the nucleus with $q_{V}=0$ at the nucleus and an increase of $q_{V}$ with the increasing distance $d$. Beyond $d=1000\km$, the model can essentially reproduce the data for the tailward region, but its agreement with the data in the sunward direction is poor. This indicates that the realistic structure of the cometary coma is indeed more complex than the idealized model of the spherical coma with constant expansion velocity.

\section{Discussion}\label{sec:dis}

\subsection{RAT alignment as a predictive theory}

For decades, the theory of grain alignment had been rather hand waving, incapable of quantitative predictions. This includes the paramagnetic alignment, which failed observations and the RAT mechanism at its initial stages of development.\footnote{The first numerical studies of the RAT alignment were inconclusive as it was somewhat a leap of faith to accept the universal mechanism on the basis of studying the alignment of a couple of selected grain shapes. Moreover, the quantitative predictions of the alignment were impossible even for those shapes as the essential processes of crossovers were not treated correctly.} However, now, with the advent of the quantitative theory of grain alignment, it is possible to make quantitative predictions about polarization. This allows polarimetry to be used to obtain reliable predictions about magnetic fields. This also allows us to predict circumstances where we expect to observe polarization (both linear and circular) arising from aligned grains. The latter two points have been explored in this paper. Some points of the present study have been mentioned/discussed earlier (see \citealt{2007JQSRT.106..225L}), but this paper is the actual quantitative study.

With grain alignment not being mysterious any more, it is getting possible to reliably relate polarimetry data and underlying magnetic fields. This should help to better understand the role of magnetic field in star formation,\footnote{Recently, a paradigm of star formation based on ambipolar diffusion has been challenged and a new model based on turbulent magnetic reconnection (\citealt{1999ApJ...517..700L}) was suggested (see \citealt*{2012ApJ...757..154L} and references therein). The predictions of the new theory based on the process that was termed "reconnection diffusion" can be tested with high resolution polarimetry data.} the contamination of microwave polarization by emission from aligned dust grain that is important for {\it Planck} mission as well as future missions. As in the years to come, the amount of data obtained with emission and extinction polarimetry data is going to dramatically increase, the theory-based interpretation of the data is getting absolutely essential.

Observational testing of the grain alignment theory is also important (\citealt{2010ApJ...720.1045A}; \citealt{2011A&A...534A..19A}; Hoang, Lazarian \& Andersson 2013). This paper contains more theoretical predictions that open ways for further observational studies of grain alignment.

\subsection{Grain alignment in the local ISM}
The LISM is an ideal location to test the different mechanisms of grain alignment thanks to its proximity. However, it is rather challenging because the degree of polarization of nearby stars induced by aligned grains is very small due to the low density of the LISM. \cite{2010MNRAS.405.2570B} found a linear increase of polarization with the increasing distance for the stars in the range $d<100\pc$, whereas \cite{2012ApJ...760..106F} showed a nearly flat feature of polarization for the stars in the range $d=5-40\pc$.

Due to its strong magnetic field and low gas density, the LISM was thought being a favored environment for the Davis-Greenstein alignment mechanism (see e.g., \citealt{Frisch:2006wc}). However, for the $\sim 0.1\mum$ grains in the LISM, we found that the rotational damping by IR emission is more efficient than by the gas bombardment, which results in the fast randomization of grain orientation compared to the Davis-Greenstein alignment timescale. Alternatively, the RAT alignment was found to be the dominant mechanism working in the LISM.

Our predictions for the starlight polarization based on RAT alignment show that the polarization increases with the increasing distance to the nearby ($d<100$ pc) stars, which is consistent with the observational data from PlanetPol. If the conclusions \cite{2012ApJ...760..106F} are true, then the flat polarization versus distance of stars for $d=5-40$ pc may be due to (i) the lack of grain alignment,
(ii) very little dust, or (iii) strongly random magnetic field in the $d>5$pc region along the sightlines observed by \cite{2012ApJ...760..106F}.

In particular, we found that the polarization curve of the nearby stars peaks at $\lambda_{\max}=0.4\mum$. This peak wavelength is much lower than the typical value ($\lambda_{\max}=0.55\mum$) of the diffuse ISM. As a result, we can test the efficiency of RAT alignment in the LISM by observing the entire polarization curve.

\subsection{Grain alignment and polarized thermal emission from accretion disks}
\cite{Tamura:1999p6500} have measured the polarization about $3\%$ of submillimeter thermal emission from the T Tauri stars. CL07 carried out detailed modeling of polarized thermal emission from aligned dust grains in an accretion disk using the RAT alignment. They predicted a level of $2\%-3\%$ for polarized thermal emission at $\lambda=100\mum$, which is comparable to the observational data by \cite{Tamura:1999p6500}. Recently, the SMA observations by \cite{2009ApJ...704.1204H} show no considerable level of polarization from the disks around two nearby stars (see Figures 2 and 3 in \citealt{2009ApJ...704.1204H}). They placed a $3\sigma$ upper limit for the total polarization of less than $1\%$. In particular, the higher resolution observations in \cite{2013AJ....145..115H} show a much lower polarization level ($\le 0.5\%$) in the disks at $100\AU$ scales. They essentially detected no polarized emission from the disks around GM Aur, MWC 480, or DG Tau.

It is worth noting that at the time of CL07's work, the understanding of RAT alignment was still incomplete. Later studies (LH07; Hoang 
\& Lazarian 2008, 2009b) showed that: (1) only a fraction $f_{\hi}$ of grains are aligned with high-$J$ attractor points, except when grains contain superparamagnetic inclusions; (2) the RAT alignment decreases with the increasing angle between the radiation anisotropy direction and the ambient magnetic field; and (3) the alignment of very large grains present in accretion disks for which the internal relaxation was negligible was weaker than the alignment of the interstellar grains. All three properties gives rise to the decrease of the polarization that was predicted by CL07.

We showed that the second parameter is the most important one. Our modeling shows that for the toroidal magnetic fields usually assumed for the T Tauri disks, dust grains in the surface layers illuminated by radiation from the central star perpendicular to the magnetic fields are weakly aligned. Grains in the disk interior mostly irradiated by the reemision of hot dust in the surface layers along the direction of rotation axis are weakly aligned because the radiation field is perpendicular to the magnetic field as well. As a result, we found that the maximum polarization is decreased substantially compared to the case in which the radiation anisotropy direction is parallel to the magnetic field. 

In realistic conditions of the accretion disks, the magnetic field is not perfectly perpendicular to the radiation direction. Therefore, we expect to see intermediate polarization of dust emission but a rather low level of polarization. Moreover, the fact that SMA observations show very low level of polarized emission reveals that the magnetic field geometry is mostly parallel to the disk plane (or toroidal) as expected. 

\subsection{Grain alignment in the Zodiacal cloud and implications for CMB studies}

A very important consequence of our study is the prediction of grain alignment in the ZC. Such a prediction is supported by a good agreement between CP predicted by our model with the observational data. Therefore, if Zodiacal dust is aligned, then the thermal emission by Zodiacal dust will be polarized. Polarized emission of Zodiacal dust would become an important component of polarized Galactic foreground that contaminates to the polarized CMB signal. Detailed modeling of polarized Zodiacal emission is necessary, and it should be taken into account within the ongoing CMB studies.

\subsection{Interplanetary magnetic field via circular polarization of Zodiacal light}
In this paper, we found that dust grains in ZC can be aligned by solar radiation with respect to the IMF. The scattering of sunlight by aligned grains in ZC appears to be a principal mechanism for producing CP of Zodiacal light. Therefore, modeling CP of Zodiacal light allows us to probe the IMF and composition of interplanetary dust.

Observations by the {\it Ulysses} spacecraft have confirmed the Parker model of the IMF but also revealed the potential existence of a vertical component (\citealt{1996JGR...101..395F}). In particular, the observational data demonstrate that the angle between the magnetic field and the ecliptic plane varies with the heliographic latitude. The angle deviation of the observed magnetic field from the Parker model ranges from $-6^{\circ}$ to $4^{\circ}$.

Our predictions for CP arising from scattering by oblate spheroidal grains that are aligned with the magnetic field show that there should exist a regular vertical component (as also pointed out in \citealt{1978A&A....69..421D}). We also found that our model with the various direction of $B_{\perp}$ is in better agreement with the observational data than the model with the regular magnetic field. This finding is reinforced by the observational data from the {\it Ulysses} spacecraft.

\cite{1978A&A....69..421D} suggested another possibility to reproduce the observed CP of Zodiacal light. Indeed, they assumed that only prolate spheroidal grains are aligned within the inner regions with $r<R_{1}$, while only oblate spheroidal grains are aligned within the anneal of $R_{1}<r<R_{2}$. Thus, the lines of sight with $\epsilon <90^{\circ}$ or $\epsilon>270^{\circ}$ go through a medium of aligned prolate spheroidal grains, and the lines of sight with $\epsilon=90^{\circ}-270^{\circ}$ go through a medium of aligned oblate spheroidal grains. Since the CP by prolate and oblate grains has opposite sign, the former gives rise to $q_{V}$ in opposite sign with the latter. Although their results could reproduce the observational data, the issue of why prolate spheroidal grains are segregated from oblate spheroidal grains is difficult to justify. One possible explanation for that is that oblate grains with larger cross section are swept away by solar radiation faster than prolate grains. However, we don't know which shape interplanetary grains look like. 

\subsection{Circular polarization and alignment mechanism in cometary comae}
CP in cometary comae is rather mysterious. The regimes in which the alignment axis happens with respect to the radiation flux failed to account for the observed CP because its alignment direction is uniform along any line of sight. While, in the ZC, the alignment with the spiral magnetic field is found to produce sufficient polarization level, this alignment type does not work in the coma because the magnetic field can not penetrate to such a deep in the coma. However, independent research testifies of the existence of electric fields throughout the coma. This electric field is likely directed in the radial direction along the gas flow. The fast precession of the grain dipole moment around the electric field makes it an alignment axis.

We perform a simple modeling of grain alignment and provide predictions for circular polarization in an idealized spherical cometary coma. Our model can reproduce the generic feature of the CP but a better fit to the observation can be obtained with a more sophisticated model of the coma. Further observational studies of linear polarization of starlight from cometary comae are necessary for testing grain alignment theory.

\subsection{Studying turbulence with aligned grains}

Polarimetry mostly deals with time independent signal. However, the cases of ZC and the cometary coma may present variable signal due to turbulence (see \citealt{2007JQSRT.106..225L}). Thus, this provides a useful way of studying turbulence on timescales larger than the precession time of angular momentum around the magnetic field, including compressible magnetic turbulence (see \citealt{1997ApJ...485..680G}; \citealt{2003MNRAS.345..325C} and references therein) in interplanetary medium and compressible multi-component turbulence in cometary coma. We believe that additional cost effective ways of studying interplanetary turbulence can help to resolve existing controversies related to the actual scaling of MHD turbulence (\citealt{2009ApJ...702.1190B}; \citealt{Beresnyak:2012ek}).

The technique of turbulence study that we are discussing here is complimentary to the techniques of in situ measurements of turbulence with spacecraft (see \citealt{2013ApJ...765...35B}) and the proposed new technique of using atomic alignment (see \citealt{2012JQSRT.113.1409Y} and references therein).

\section{Summary}\label{sec:sum}
With the advent of the quantitative theory of RAT alignment, it became possible to make predictions for expected linear and circular polarization by aligned grains. The principal results of our paper can be summarized as follows.

\begin{itemize}

\item[1.] We studied the alignment of grains in the LISM and showed that the alignment by RATs is dominant. The linear polarization of nearby stars predicted by the RAT alignment was found in good agreement with the observational data, which exhibit the increase of polarization with the distance to the stars.

\item[2.] The problem of grain alignment in the T Tauri disks was revisited accounting for the dependence of grain alignment on the angle between the radiation anisotropy direction and the magnetic field. We show that for the disk with a toroidal magnetic field, grains are weakly aligned because the anisotropy direction is nearly perpendicular to the magnetic field. The result can explain the very low degree of submillimeter polarization recently observed from T Tauri disks.

\item[3.] We found that grains in the ZC can be aligned by RATs induced by solar radiation. The predictions for CP of Zodiacal light by aligned grains seem to be consistent with the existing CP data and plausible configuration of the IMF. We suggest observers to perform new studies of circular and linear polarization from the ZC. This can improve our knowledge of the magnetic field structure in the Solar system and this study is necessary for understanding the contribution of polarized Zodiacal emission to the CMB experiments. 

\item[4.] We studied grain alignment and calculated the CP of scattered light from a cometary coma. For this special environment, we suggest a new alignment mechanism based on the action of RATs over grains precessing around electric fields. Using the proposed alignment mechanism, we can reproduce the systematic changes of CP across the coma as reported by observers.

\end{itemize}

\section*{Acknowledgments}
We thank the anonymous referee for useful comments and suggestions that significantly improved our paper. A.L. acknowledge the financial support of NASA grant NNX11AD32G and the Center for Magnetic Self-Organization. The initial calculations for the Zodiacal light were started with the help of an undergraduate REU student Anna Boehle whom we would like to thank for her enthusiasm and assistance.

\appendix

\section{Radiative torque (RATs) and RAT Efficiency}\label{apdx:a}

Similar to \cite{2007MNRAS.378..910L}, in order to make an easy relation of our results to those in earlier works, wherever it is possible, we preserve notations adopted in DW97. Mean radiative torque efficiency over wavelengths, $\overline{\bQ(\Theta,\beta,\Phi)}$ is defined as
\bea
\overline{\bQ}=\frac{\int \bQ_{\lambda}u_{\lambda}d\lambda}
{\int u_{\lambda}d\lambda} \label{eq:a01}, 
\ena 
where $u_{\lambda}$ is the energy density (see \citealt{1983A&A...128..212M}), and
$\bQ_{\lambda}$ is the RAT efficiency corresponding to wavelength $\lambda$. RAT from the anisotropic component of radiation is defined by 
\bea 
\bGamma_{\rad}=\frac{u_{\rad}a_{\eff}^{2}\overline{\lambda}}{2}\gamma\overline{\bQ},\label{eq:a02}
\ena 
where $\gamma$ is the anisotropy degree of radiation, $a_{\eff}$ is the effective grain size (see DW96; Paper I), $u_{\rad}$ and $\overline{\lambda}$ are the energy density and mean wavelength of radiation field. The latter are respectively given by
\bea
u_{\rad}&=&\int u_{\lambda}d\lambda,\\
\overline{\lambda}&=&\frac{\int \lambda u_{\lambda}
d\lambda}{u_{\rad}}.\label{eq:a03}
\ena

The specific energy density of a radiation field with the intensity
$I_{\lambda}$
is defined as
\bea
u_{\lambda}(\Omega)=\frac{I_{\lambda}}{c}.
\ena
Integrating over all solid angles we obtain
\bea
u_{\lambda}=\int u_{\lambda}(\Omega)d\Omega=\frac{1}{c}\int I_{\lambda}d\Omega.\label{eq:ulam}
\ena
Defining the mean intensity over the solid angle
\bea
J_{\lambda}=\frac{1}{4\pi}\int I_{\lambda}d\Omega,
\ena
Equation (\ref{eq:ulam}) can be rewritten as
\bea
u_{\lambda}=\frac{4\pi}{c}J_{\lambda}
\ena

For an isotropic radiation field from a star of temperature $T_{\star}$ so that $I_{\lambda}=J_{\lambda}\equiv B_{\lambda}$, $u_{\lambda}$
becomes
\bea
u_{\lambda}=\frac{4\pi}{c}B(\lambda,T_{\star}),\label{eq:a05}
\ena
where
\bea
B(\lambda,T_{\star})=\frac{2hc^{2}}{\lambda^{5}}\frac{1}{\exp\left(hc/\lambda
k_{\B}T_{\star}\right)-1},\label{eq:a06}
\ena

\section{Extinction and Polarization Cross Section}\label{sec26}
\subsection{Extinction Cross Section}
Let us consider a spheroid grain with the symmetry axis $\ba_{1}$. A perfectly polarized electromagnetic wave with the electric field vector $\bE$ propagates along the $z$-axis,
which is perpendicular to the symmetry axis. Let $C_{\ext}(\bE\perp \ba)$ and $C_{\ext}(\bE\| \ba)$ be the extinction of the radiation for the cases in which the electric field vector
is parallel and perpendicular to the grain symmetry axis, respectively.

For simplification, we denote these extinction cross section by $C_{\perp}$ and $C_{\|}$.
For the general case in which $\bE$ makes an angle $\theta$ with the symmetry axis, the extinction cross section becomes
\bea
C_{\ext}=\cos^{2}\theta C_{\|}+\sin^{2}\theta C_{\perp}.\label{eq:Cext0}
\ena

For a randomly oriented grain, one can compute the total extinction cross section by integrating Eq. (\ref{eq:Cext0}) over the isotropic distribution of $\theta$, i.e., $f_{\rm iso} d\theta \sim \sin\theta d\theta$. As a result,
\bea
C_{\ext}=\frac{1}{3}\left(2C_{\perp}+C_{\|}\right).\label{eq:Cext}
\ena

The polarization efficiency is defined as
\bea
C_{\pol}&=&C_{\perp}-C_{\|},\label{eq:Cpol0}\\
C_{\pol}&=&\frac{1}{2}\left(C_{\|}-C_{\perp}\right),
\ena
for oblate and prolate spheroidal grains, respectively.

\subsection{Polarization Cross Section}
Consider an observer's reference system, which is defined by the line of sight directed along the $z$-axis, the projection of magnetic field on the sky plane denoted by the $y$-axis, and the third axis is perpendicular to the $yz$ plane, namely $x$-axis. Thus, $\bB$ lies in the $yz$ plane and makes a so-called angle $\xi$ with the $y$-axis.

By transforming the grain coordinate system to the observer coordinate system and taking corresponding weights, we obtain
\bea
C_{x}&=&C_{\perp}-\frac{C_{\pol}}{2}\sin^{2}\beta,\\
C_{y}&=&C_{\perp}-\frac{C_{\pol}}{2}(2\cos^{2}\beta\cos^{2}\xi+\sin^{2}\beta\sin^{2}\xi),
\ena
where the perfect internal alignment of grain axes with the angular momentum has been assumed.

The polarization cross section then becomes
\bea
C_{x}-C_{y}=C_{\pol}\frac{\left(3\cos^{2}\beta-1\right)}{2}\cos^{2}\xi.\label{eq:Cpol}
\ena
Taking the average of $C_{x}-C_{y}$ over the distribution of the alignment angle $\beta$, the above equation can be rewritten as
\bea
C_{x}-C_{y}=C_{\pol}\langle Q_{J}\rangle \cos^{2}\xi,\label{eq:Cx-Cy}
\ena
where
\bea
Q_{J}=\frac{\left(3\cos^{2}\beta-1\right)}{2}
\ena
is the degree of alignment of the grain angular momentum with the ambient magnetic field. 

When the internal alignment is not perfect, following the similar procedure, we obtain
\bea
C_{x}-C_{y}=C_{\pol}\langle Q_{J}Q_{X} \rangle \cos^{2}\xi \equiv
C_{\pol}R\cos^{2}\xi,\label{eq:Cpoln}
\ena
where $R=\langle Q_{J}Q_{X}\rangle$ is the Rayleigh reduction factor (see also \citealt{1999MNRAS.305..615R}).

In the above equation, $Q_{X}$ is the degree of internal alignment of grain axes with angular momentum. In the case of fast internal relaxation, the angle between $\ma_{1}$ and $\bJ$ fluctuates so fast and is described by the LTE distribution $f_{\LTE}(\theta,J)$. The internal degree of alignment $Q_{X}$ is defined as
\bea
Q_{X}(J)=\int f_{\rm LTE}(\theta,J) \sin\theta d\theta.\label{eq:QX}
\ena
For suprathermal rotation, i.e., very large $J\gg J_{\th}$, then $Q_{X}(J)\rightarrow 1$.

For the case of perpendicular magnetic field, i.e., $\Bv$ lies on the sky plane, Equation (\ref{eq:Cpol}) simply becomes $C_{x}-C_{y}=C_{\pol}R\approx C_{\pol}\langle Q_{J}\rangle \langle Q_{X}\rangle$.

The total extinction coefficient is then
\bea
C_{\ext}=\frac{C_{x}+C_{y}}{2}=\frac{2C_{\perp}+C_{\|}}{3}-\frac{\Phi
C_{\pol}}{6}
\left(3-\frac{2}{\cos^{2}\xi}\right).~~~
\ena
The first term is much larger than the second one, so the extinction coefficient can be approximated to Equation (\ref{eq:Cext}).

The wavelength dependence of optical depth $\tau_{\lambda}$ along a line of sight is obtained by integrating Equation (\ref{eq:Cext}). The interstellar extinction is then given by 
$A_{\lambda}=\left(2.5/\ln10\right)\tau_{\lambda}=1.086\tau_{\lambda}$.

The extinction coefficients $C_{\|}$ and $C_{\perp}$ for oblate spheroidal grains with the ratio of semiaxes $s=b/a=0.5$ are computed using the DDSCAT code (\citealt{1994JOSAA..11.1491D}). 

\section{Circular polarization Cross Section due to scattering by aligned grains}\label{apdx:b}
In the following, we derive the expression of circular polarization \ref{eq:V}. We consider a grain model consisting of dipole moments (\citealt{1973ApJ...186..705P}; \citealt{1973MNRAS.162..367B} \citealt{1988ApJ...333..848D}). The scattering problem consists of the excitation of electric field of incident light on the electric dipoles, resulting in the oscillation of dipoles. The oscillation of electric dipoles reemit radiations in every direction, but the direction perpendicular to the oscillation (acceleration) direction is strongest. During the oscillation, some damping process within the grain can result in the damping of the oscillation, converting oscillation energy into heat, which decreases the amplitude of reemitting light compared to the amplitude of incident light. Then, one can say that some absorption of radiation energy occurs. Thus, the scattering and absorption is not independent.

In addition to the elastic scattering (i.e., the frequency of radiation does not change in the elastic scattering), the phase of electric field vectors may vary by the scattering, which induces the circular polarization.
 
The properties of radiation is described in general by the four Stokes parameters denoted by a vector F=(I, Q, U,V). The scattering light is then given by
\bea
F^{\rm sca}_{i}=\frac{1}{k^{2}r^{2}}S_{ij} F^{\rm inc}_{j},\label{eq:Scamatrix}
\ena
where $S_{ij}$ is a scattering matrix of $4\times 4$, $k$ is the wave vector and $r$ is the distance from the grain to the observer (see \citealt{1983asls.book.....B}, p.65). The scattering matrix $S$ represents the complete properties of the scattering process by a single grain. The Stokes parameters of light scattered by an ensemble of grains are the sum of the Stokes parameters of light scattered by each grain. Thus, one can describes the scattering by the ensemble by a net scattering matrix in which the elements are the sum of the elements of the scattering matrix from each grain, i.e., $S_{ij}^{\rm net}=\sum_{i=1}^{N_{\rm gr}}S_{ij}^{i}$.

During the scattering, in addition to the change of the radiation energy, the phase of electric field vector also changes, which results in the circular polarization. From Equation (\ref{eq:Scamatrix}), the circular polarization is described by the elements $S_{41}$ if the incident light is unpolarized, i.e., $Q, U$ and $V$ are zero:
\bea
V_{\rm sca}=\frac{1}{k^{2}r^{2}}S_{41} I_{\rm inc},\label{eq:Vsca}
\ena

The element $S_{41}$ depends on the refractive index of the dust grain, which is given by
\bea
S_{41}\propto {\rm Im}(\alpha_{1}\alpha_{3}^{*}) \sin\phi\sin\psi\cos\psi\sin\theta,
\ena
where $\alpha$ is the complex polarizability characterizing the reaction of material with the electric field (i.e., the instantaneous dipole $j$ with $\bP_{j}=\alpha_{j} \bE_{j}$), $\theta$ is the scattering angle, $\phi$ and $\psi$ describes the orientation of the grain symmetry axis (also the magnetic field direction) in the lab frame (see \citealt{2000MNRAS.314..123G}). Indeed, $\psi$ is the angle between the grain symmetry axis and the scattering plane.

Let $\sigma_{V}$ the circular polarization cross section, then one can write
\bea
\sigma_{V}&\propto& -{\rm Im} (\alpha_{1}\alpha_{3}^{*})\nonumber\\
&=&\left[Re(\alpha_{1})Im(\alpha_{3})-Im(\alpha_{1})Re(\alpha_{3})\right].
\ena

For axisymmetric grain, we have $\alpha_{1}=\alpha_{\|}$ and $\alpha_{2}=\alpha_{3}=\alpha_{\perp}$. Thus, the above equation becomes
\bea
\sigma_{V}\propto \left[Re(\alpha_{\|})Im(\alpha_{\perp})-Im(\alpha_{\|})Re(\alpha_{\perp})\right]
\ena
identical to Equation (\ref{eq:sigmaV}).

For the limit of small grains, $x=ak=2\pi a/\lambda\ll 1$, using the complex refractive index, Equation (\ref{eq:sigmaV}) can be rewritten as
\bea
\sigma_{V}&=&k^4\left[Re(\alpha_{\|}) Im(\alpha_{\bot})-Im(\alpha_{\|})Re(\alpha_{\bot})\right],\\
&=&\left(\frac{2\pi}{\lambda}\right)^{4}\left[Re(\alpha_{\|}) Im(\alpha_{\bot})-Im(\alpha_{\|})Re(\alpha_{\bot})\right].
\ena

For isotropic material or sphere, the polarizability $\alpha$ is a scalar and given by
\bea
\alpha=a^{3}\left(\frac{m^{2}-1}{m^{2}+2}\right),
\ena
where $m$ is the reflective index of the medium.

For anisotropic material, the polarizability $\alpha$ becomes a tensor with the diagonal elements
\bea
\alpha_{j}=\frac{V}{4\pi}\left(\frac{1}{L_{j}+(m^{2}-1)^{-1}}\right),\label{eq:alpha}
\ena
where $V$ is the grain volume, $L_{j}$ are the shape factors (see \citealt{1957lssp.book.....V}, p. 71). 

One can see that the circular polarization requires the imaginary part of the reflective index, indicating that the circular polarization is induced by the absorption of the light.

\bibliography{ms.bbl}

\end{document}